\documentclass[letterpaper]{article}

\usepackage[sort&compress]{natbib}
\usepackage{url}

\newcommand{\mfour}{\mathcal{M}_4}
\newcommand{\actd}{{D_{\!A}}}
\newcommand{\actdiff}{\mathrm{Diff}_A \, \mfour}
\newcommand{\passdiff}{\mathrm{Diff}_P \, \mfour}
\newcommand{\geom}{\mathrm{Geom} \, \mfour}
\newcommand{\riem}{\mathrm{Riem} \, \mfour}
\newcommand{\eqref}[1]{(\ref{#1})}

\title{Ephemeral point-events: is there a last remnant of physical
objectivity?\footnote{This Essay is dedicated with warm affection to
Roberto Torretti on the occasion of his $70^{\mathrm{th}}$
Birthday. Most of the technical developments that underlay this work
were introduced by \citet{Lusannapauri00}. Some of this material was
also discussed at the international workshop \emph{General covariance
and the quantum: where do we stand?}, held at the University of Parma
on June 21--23, 2001. We are deeply indebted to Luca Lusanna for a
long series of enlightening discussions about the canonical reduction
of general relativity and about the Bergmann--Komar intrinsic coordinates.}}

\author{Massimo Pauri$^1$ and Michele Vallisneri$^2$ \\[10pt]
{\small ${}^1$ Dipartimento di Fisica, Universit\`a di Parma, 43100 Parma, Italy} \\
{\small ${}^2$ Theoretical Astrophysics 130-33, Caltech, Pasadena CA 91125, USA}}

\date{Jan 21, 2002 (last revised Oct 16, 2002)}

\begin{document}

\maketitle

\begin{abstract}
For the past two decades, Einstein's Hole Argument (which deals with
the apparent indeterminateness of general relativity due to the
general covariance of the field equations) and its resolution in terms
of \emph{Leibniz equivalence} (the statement that Riemannian
geometries related by active diffeomorphisms represent the same
physical solution) have been the starting point for a lively
philosophical debate on the objectivity of the point-events of
space-time.  It seems that Leibniz equivalence makes it impossible to
consider the points of the space-time manifold as \emph{physically
individuated} without recourse to dynamical individuating fields.
Various authors have posited that the metric field itself can be used
in this way, but nobody so far has considered the problem of
explicitly distilling the \emph{metrical fingerprint} of point-events
from the gauge-dependent components of the metric field.  Working in
the Hamiltonian formulation of general relativity, and building on the
results of \citet{Lusannapauri00}, we show how Bergmann and Komar's
\emph{intrinsic pseudo-coordinates} (based on the value of curvature
invariants) can be used to provide a physical individuation of
point-events in terms of the true degrees of freedom (the \emph{Dirac
observables}) of the gravitational field, and we suggest how this
conceptual individuation could in principle be implemented with a
well-defined empirical procedure.  We argue from these results that
point-events retain a significant kind of physical objectivity.
\end{abstract}

\vspace{2cm}

\section{Introduction: Einstein, the Hole
Argument, and the physical individuation of point-events}

General relativity owes much of its mathematical beauty to its
formulation in terms of the theory of pseudo--Riemannian manifolds.
This beauty, however, carries a curse: at the
mathematical level, even a \emph{naked} manifold has well-defined
points distinguishable in terms of coordinates, but in physics it is a
widely held assumption that points can be distinguished only by
the values of physical fields or as the positions of physical
objects, including measuring devices. Any attempt to take the bare
points seriously leads to well-known puzzles and quandaries.

Possibly the first puzzle of this kind (the Hole Argument, or
\emph{Lochbetrachtung}) crossed Albert Einstein's path repeatedly
between 1913 and 1915. These were years of alternating joy and
distress for Einstein, as he set out to create a theory of gravitation
based on the guiding principle of general covariance, failed to do so,
used the Hole Argument to convince himself that general covariance was
physically inconsistent, formulated the short-lived
Einstein--Grossmann (\emph{Entwurf}) theory, and finally returned to
his original conviction, having come, through the Hole Argument, to
his explanation of the physical meaning of general covariance.
Roberto Torretti wrote a beautiful account of Einstein's woes and
triumphs in his masterly treatise \emph{Relativity and Geometry}
\citep{Torretti00}, and more about this story can be found in John
Norton's contribution to this very volume, as well as in many other
papers by Norton
(\citeyear{Norton00,Norton01,Norton02,Norton04,Norton05,Norton06,Norton07};
see also Howard and Norton, \citeyear{Howardnorton00}) and by John
Stachel
\citeyearpar{Stachel00,Stachel01,Stachel02,Stachel03,Stachel04}.

Einstein's ``triumph'' [to use Norton's wording \citeyearpar{Norton03}] over
the Hole Argument and ``over the space-time coordinate systems'' came
only after he adopted a very idealized model of physical measurement
where all possible observations reduce to the intersections of the
worldlines of observers, measuring instruments, and measured physical
objects (\emph{point-coincidence argument}). In Einstein's own
words \citeyearpar{Einstein00}:
\begin{quotation}
\noindent That the requirement of general covariance, \emph{which
takes away from space and time the last remnant of physical
objectivity}, is a natural one, will be seen from the following
reflexion. All our space-time verifications invariably amount to a
determination of space-time coincidences. If, for example, events
consisted merely in the motion of material points, then ultimately
nothing would be observable but the meetings of two or more of
these points. Moreover, the results of our measurings are nothing
but verifications of such meetings of the material points of our
measuring instruments with other material points, coincidences
between the hands of a clock and points on the clock dial, and
observed point-events happening at the same place at the same
time. The introduction of a system of reference serves no other
purpose than to facilitate the description of the totality of such
coincidences.
\end{quotation}

\noindent The Hole Argument received new life with John Stachel's seminal paper
\citeyearpar{Stachel00}, which raised a rich philosophical
debate\footnote{See
\citet{Earmannorton00,Norton00,Norton01,Norton02,Norton03,Norton04,
Norton05,Norton06,Norton07,Butterfield00,Butterfield01,Butterfield02,
Butterfield03,Fine00,Belot00,Belot01,Brighouse00,Hofer00,Hofer01,
Hofercartwright00,Chuang00,Chuang01,Leeds00,Maudlin00,Maudlin01,
Rynasiewicz00,Rynasiewicz01,Rynasiewicz02,Teller00,Wilson00,
Disalle00,Saunders00,Bartels00,Stein00}.  In this paper we shall make
no attempt to analyze or survey this discussion, not least because
we believe that some debaters occasionally overstep the philosophical latitude allowed
by the very structure of general relativity. Instead, we shall recall
only the major points that can be seen as a premise to our
discussion.} that is still alive today.  Soon it became widely
recognized that the Hole Argument was intimately tied with our
conceptions of space and time, at least as they are represented by the
mathematical models of general of relativity.

Of course, it is to philosophical preferences that we must defer
the judgment on the ontological status of the notions that are introduced
in physical theories to describe Nature; and this is especially
true for the conditions that decide in favor of a literal or
nonliteral interpretation of theoretical structures.  So we shall
not be concerned here with the metaphysical issue of the reality
or nature of space-time, let alone of the \emph{Raum} of our
experience. We agree with Michael Friedman when he argues that the
Hole Argument leaves an unsolved problem  about the
characterization of intrinsic space-time structure, rather than an
ontological question about the existence of space-time [``avoiding
quantification over `bare' points \ldots\ appears to be a non-trivial
mathematical problem'' \citep{Friedman02}].

In this paper we offer our contribution to the clarification of this
non-trivial problem. More precisely, we investigate the relation
between the physical meaning of spatio-temporal localization and the
unavoidable use of \emph{arbitrary} coordinate systems in the practice
of general relativity. Thus, we explore the limits on the objectivity
of space-time that are imposed by the mathematical representation of
spatio-temporal structure, in conjunction with
the requirements of the empirical foundation of general relativity.

\subsection{The Hole Argument}

In its modern version, the Hole Argument runs as follows. Consider a
general-relativistic space-time, as specified by a four-dimensional
mathematical manifold $\mathcal{M}_4$ and by a metrical tensor field $g$,
which represents at the same time the chrono-geometrical structure of
space-time and the potential for the gravitational field. The metric
$g$ is a solution of the generally-covariant Einstein equations. If
any nongravitational physical fields are present, they are represented
by tensor fields that are also dynamical fields, and that appear as
sources in the Einstein equations.

Now assume that $\mathcal{M}_4$ contains a \emph{hole}
$\mathcal{H}$: that is, an open region where all the
nongravitational fields are null. On $\mathcal{M}_4$ we can prescribe an
\emph{active}\footnote{\label{note:trick}As originally formulated
by \citet{Einstein01}, the Hole Argument does not rely on
the effects of active diffeomorphisms in the modern geometrical
sense, but rather on the following procedure. After taking a
coordinate transformation $\hat{\xi}^\mu = f^\mu(\xi^\nu)$, we
obtain the transformed metric $\hat{g}_{\sigma
\rho}(\hat{\xi}^\mu)$, and then we consider the object
$\hat{g}_{\sigma \rho}(\xi^\mu)$ defined by transferring the
functional dependence of $\hat{g}_{\sigma \rho}(\hat{\xi}^\mu)$ to
the \emph{old} coordinates $x^\mu$. This is akin to obtaining an
active diffeomorphism as the \emph{dual} of a passive
transformation.} diffeomorphism $\actd$
\citep{Wald00,Stachel02,Norton00} that remaps the
points inside $\mathcal{H}$, but blends smoothly into the identity
map outside $\mathcal{H}$ and on the boundary. Because the
Einstein equations are generally covariant, if $g$ is one of their
solutions, so is the \emph{drag-along} field $g' = \actd g$. By
construction, for any point $x \in \mathcal{H}$ we have
(geometrically) $g'(\actd x)=g(x)$, but of course $g'(x) \neq
g(x)$ (also geometrically).

What is the correct interpretation of the new field $g'$? Clearly, the
transformation entails an \emph{active redistribution of the metric
over the points of the manifold}, so the crucial question is whether, to
what extent, and how the points of the manifold are primarily
\emph{individuated}.\footnote{Consistently with our program, we
shall not get involved in the deep philosophical issue of the
\emph{individuation} of entities in general. Throughout this essay, our
notion of individuation will be deliberately restricted to the meaning that it
can have at the mathematical level and, above all, within the
conceptual context of a physical theory.} In the mathematical
literature about topological spaces, it is always implicitly
assumed that the entities of the set can be distinguished and
considered separately (provided the Hausdorff conditions are
satisfied), otherwise one could not even talk about point mappings or
homeomorphisms. It is well known, however, that the points of a
homogeneous space cannot have any intrinsic \emph{individuality}. As
Hermann Weyl \citeyearpar{Weyl00} put it:
\begin{quotation}
\noindent There is no distinguishing objective property by which
one could tell apart one point from all others in a homogeneous
space: at this level, fixation of a point is possible only by
a \emph{demonstrative act} as indicated by terms like ``this'' and
``there.''
\end{quotation}

\noindent Quite aside from the phenomenological stance implicit in
Weyl's words,\footnote{One could contemplate stripping the argument
from its phenomenological flavor by asserting that, after all, the
demonstrative act also establishes an empirical coincidence. This view
is taken, for instance, by Moritz Schlick \citeyearpar{Schlick00}, who
writes: ``In order to fix a point in space, one must somehow directly
or indirectly, \emph{point to it} \ldots\ that is, one establishes a
spatio-temporal coincidence of two otherwise [already] \emph{separate}
elements.''} there is only one way to individuate points at the
mathematical level that we are considering: namely by
coordinatization, which transfers the individuality of $n$-tuples of
real numbers to the elements of the topological set. Therefore, all
the relevant transformations (including \emph{active} diffeomorphisms)
operated on the manifold $\mfour$, even if viewed in purely
geometrical terms, \emph{must} be constructible in terms of coordinate
transformations (see for instance note \ref{note:trick}). So we have
necessarily crossed from the domain of \emph{geometry} to
\emph{algebra}, and we can justify our use of the symbol $x$ to
denote a point of the manifold, as mathematically individuated by
the chosen coordinates.

Let us go back to the effect of this \emph{primary} individuation
of manifold points. If we now think of the points of $\mathcal{H}$
as already \emph{independently individuated} spatio-temporal
\emph{physical events} even before the metric is defined, then $g$
and $g'$ must be regarded as \emph{physically distinct} solutions
of the Einstein equations (after all, $g'(x) \neq g(x)$ at the
\emph{same} point $x$). This is a devastating conclusion for the
causality, or better, \emph{determinateness}\footnote{We prefer to
avoid the term \emph{determinism}, because we believe that its
metaphysical flavor tends to overstate the issue at hand. This is
especially true if \emph{determinism} is taken in opposition to
\emph{indeterminism}, which is not mere absence of
\emph{determinism}.} of the theory, because it implies that, even
after we completely specify a physical solution for the
gravitational and nongravitational fields outside the hole (for
example, on a Cauchy surface for the initial value problem), we
are still unable to predict uniquely the physical solution within
the hole.  Clearly, if general relativity has to make any sense as
a physical theory, there must be a way out of this foundational
quandary, \emph{independently of any philosophical consideration}.

In the modern understanding, the most widely embraced escape from the
strictures of the Hole Argument (which is essentially an update to
current mathematical terms of the naive solution adopted by Einstein),
is to \emph{deny that diffeomorphically related mathematical solutions
represent physically distinct solutions}. With this assumption,
\emph{an entire equivalence class of diffeomorphically related
mathematical solutions represents only one physical
solution}.\footnote{Of course, taken at face value this statement
could be misinterpreted as the naive (and physically vacuous)
assertion that metric tensors that have different descriptions in
different coordinate systems are geometrically the same tensor
(invariance with respect to \emph{passive} diffeomorphisms
$\passdiff$). To formulate the Hole Argument, however, we have used
\emph{active} diffeomorphisms: although, as said before, these are generated by the
drag-along of coordinate systems, they have the effect that the metric
tensors $g$ and $\actd g$ become \emph{geometrically different} at
each point $x \in \mathcal{H}$.}  This statement has come to be called
[after Earman and Norton \citeyearpar{Earmannorton00}] \emph{Leibniz
equivalence}.

It should be clear from the beginning that this is an allusion to a
\emph{new Leibniz} adapted to the modern context of general
relativity. Apart from the structural analogy, modern Leibnizian
arguments proceed without any reference to the metaphysical premises
of Leibniz's historical arguments.\footnote{More aptly, Friedman calls
this Leibniz, stripped of his metaphysical assumptions, the
\emph{Leibniz of the positivists} (Friedman, \citeyear[p.\
219]{Friedman00}; see also Friedman, \citeyear{Friedman01}). A
penetrating analysis of the \emph{old} Leibniz versus the \emph{new}
one can be found in \citet{Earman00}.} The same should be said of the
Newtonian arguments that underlie the modern version of
\emph{substantivalism} (see more below). Rynasiewicz
\citeyearpar{Rynasiewicz03} has properly remarked that, as it is often
portrayed in twentieth-century philosophical literature, even the
opposition between \emph{substantivalism} and \emph{relationism}
amounts to a historical misrepresentation of the classical
Newton--Leibniz controversy [see also \citet{Dorato00}]. This is not
irrelevant to the present considerations, for we find it rather
arbitrary to transcribe Newtonian absolutism (or at least part of it) into
the so-called \emph{manifold substantivalism}, no less than to
assert that general relativity is a \emph{relational} theory in an
allegedly Leibnizian sense. As emphasized by Rynasiewicz, the
crucial point is that the historical debate presupposed a clear-cut
distinction between \emph{matter} and \emph{space}, or between
\emph{content} and \emph{container}; but by now these distinctions have been
blurred by the emergence of the so-called \emph{electromagnetic view
of nature} in the late nineteenth century [for a detailed
model-theoretical discussion of this point see also \citet{Friedman00}].

Still, although some might argue [as do Earman and Norton
\citeyearpar{Earmannorton00}] that the metric tensor, \emph{qua}
physical field, cannot be regarded as the \emph{container} of other
physical fields, we argue that the metric field has ontological
priority over all other fields. This preeminence has various reasons
\citep{Pauri00}, but the most important is that the metric field tells
all other fields how to move causally. We also agree with Friedman
\citeyearpar{Friedman00} that, in agreement with the
general-relativistic practice of not counting the gravitational energy
induced by the metric as a component of the total energy, we should
regard the manifold, endowed with its metric, as space-time; and leave
the task of representing matter to the stress-energy tensor. Because
of this priority, beside the fact that the Hole \emph{is} pure
gravitational field, we maintain, unlike other authors [see for
example \citet{Rovelli00,Rovelli01,Rovelli02}], that the issue of the
individuation of points of the manifold as \emph{physical}
point-events\footnote{There is an unfortunate ambiguity in the usage
of the term \emph{space-time points} in the literature: sometimes it
refers to element of the mathematical structure that is the first
layer of the space-time model, sometimes to the points interpreted as
\emph{physical} events: we will adopt the term \emph{point-event} in
the latter sense and simply \emph{point} in the former.} should be
discussed primarily in the context of the vacuum gravitational field,
without any recourse to nongravitational entities, except perhaps at the
operational level.  In this paper we shall indeed adopt this choice.

Stachel
\citeyearpar{Stachel00,Stachel01,Stachel02,Stachel03,Stachel04} has
given a very enlightening analysis of the meaning of general
covariance and of its relations with the Hole Argument, expounding the
conceptual consequences of Einstein's acceptance of modern Leibniz
equivalence through the point-coincidence argument. Stachel stresses
that asserting that $g$ and $\actd g$ represent \emph{one and the same
gravitational field} is to imply that \emph{the mathematical
individuation of the points of the differentiable manifold by their
coordinates has no physical content until a metric tensor is
specified}. In particular, coordinates lose any \emph{physical}
significance whatsoever \citep{Norton03}. Furthermore, as Stachel
emphasizes, if $g$ and $\actd g$ must represent the same gravitational
field, they cannot be physically distinguishable in any way. So when
we act on $g$ with $\actd$ to create the drag-along field $\actd g$,
no element of physical significance can be left behind: in
particular, nothing that could identify a point $x$ of the manifold as
the \emph{same} point of space-time for both $g$ and $\actd
g$. Instead, when $x$ is mapped onto $x' = \actd x$, it brings over
its identity, as specified by $g'(x')=g(x)$.

This conclusion leads Stachel to the conviction that space-time
points \emph{must} be \emph{physically} individuated \emph{before}
space-time itself acquires a physical bearing, and that the metric
plays in fact the role of \emph{individuating field}. What is
more, in practice even the topology of the underlying manifold
cannot be introduced independently of the specific form of the
metric tensor, a circumstance that makes it even more implausible
to interpret the mere topological manifold as \emph{substantival
space-time} (\emph{manifold substantivalism}).

Finally, it is essential to note, once again with Stachel, that simply
because a theory has generally covariant equations, it does not follow
that the points of the underlying manifold must lack any kind of physical
individuation. Indeed, what really matters is that there can be no
\emph{nondynamical individuating field} that is specified
\emph{independently} of the dynamical fields, and in particular
independently of the metric. If this was the case, a
\emph{relative} drag-along of the metric with respect to the
(supposedly) individuating field would be physically significant and
would generate an inescapable Hole problem. Thus, the absence of any
nondynamical individuating field, as well as of any dynamical
individuating field independent of the metric, is the crucial feature
of the purely gravitational solutions of general relativity.

After a brief detour into the main themes of the philosophical debate
on the Hole, we shall come back to Leibniz equivalence and argue that
it bears little relation to the determinateness of general
relativity, and that instead it amounts to the recognition that the
mathematical representation of space-time contains superfluous
structure, which must be isolated.

\subsection{The philosophical debate on the Hole}

The modern \emph{substantivalist} position\footnote{See
\citet{Butterfield00,Butterfield01,Butterfield02,
Butterfield03,Disalle00,Fine00,Hofer00,Hofer01,Hofercartwright00,Maudlin00,
Maudlin01,Bartels00,Stein00,Teller00}.} is a statement of
spatio-temporal realism: its adherents claim that the
individual \emph{points} of the manifold, for a given solution of
the Einstein equations, represent \emph{directly} the physical
points of space-time, as they would occur in the actual or in some
possible world.

Of course, as we have already emphasized, if we do assume that the
points possess an individual existence of their own, then the
rearrangement of the metric field against their background, as
envisaged in the Hole Argument, would produce a true change in the
physical state of space-time.  For this reason, according to Earman
and Norton \citeyearpar{Earmannorton00}, substantivalism can be accused of
turning general relativity into an \emph{indeterministic} theory: if
diffeomorphically related metric fields represent different physical
states, then any prescription of initial data (outside the hole) would
fail to determine a corresponding solution of the Einstein equations
(inside the hole), because many are equally possible. Earman and Norton's
intention is to confront the substantivalists with a dire dilemma:
accept \emph{indeterminism}, or abandon \emph{substantivalism}.

There have been various attempts in the substantivalist camp to
counter this threat of \emph{indeterminism}. For example,
Butterfield \citeyearpar{Butterfield00,Butterfield01,Butterfield02,
Butterfield03} portrays diffeomorphic models as different possible
worlds and invokes counterpart theory to argue that at most one
can represent an actual space-time.  Maudlin
\citeyearpar{Maudlin00,Maudlin01} claims that a space-time can be
properly represented by at most one of two diffeomorphically
related solutions of Einstein's equations, because each space-time
point carries metrical properties \emph{essentially}, so these
properties are \emph{names} in the Kripkean sense of \emph{rigid
designators}: within a class of diffeomorphic models, only one
specimen can represent a possible world, because a world in
which a point bears metrical properties other than the ones it
actually bears would be an impossible world.

Bartels \citeyearpar{Bartels00} objects to Maudlin that ``with respect
to the concrete spots of the metrical field in our world one can
reasonably say that their metrical properties could not be otherwise
than they actually are \ldots\ But to say the same with respect to
manifold points in a model is highly problematic, because
diffeomorphisms obviously generate permissible models in which the
same manifold points bear different metrical properties.'' Bartels
then proposes to take a whole equivalence class of diffeomorphic image
points of a point $p$ as the representation of one and the same
possible space-time point, because all the diffeomorphic image points
of a certain point $p$ in a model bear the \emph{same individuating
metrical fingerprint}. Yet, independently of any philosophical
preference, this suggestion is technically not viable; for, lacking any
specific definition of such equivalence classes, it could even happen
that an equivalence class, which is supposed to represent a \emph{real
point}, actually covers all points of the manifold.  It seems
therefore that the \emph{essentialist} recourse to metrical
fingerprints as an escape from the Hole Argument is doomed to fail,
unless it is possible to give a consistent mathematical definition of
\emph{metrical fingerprint}.
Even then, we still believe that it is necessary to accept \emph{Leibniz
equivalence}, at least as a starting point. At the end of our analysis,
it should be apparent that the specific structure of the
individuating \emph{metrical fingerprint} leaves no room
to sidestep the Hole Argument with any essentialist interpretation of point-events.

Let us now have a look at Roberto Torretti's reaction to some of
these positions, and to the Hole Argument in general. In his
recent book \emph{The Philosophy of Physics} \citep{Torretti01},
Torretti argues that ``the [Hole] argument forgets the fact, so
clearly set forth by Newton, that points in a structured manifold
have no individuality apart from their structural relations.'' He
then quotes Newton's \emph{De Gravitatione}
\citep{HallandHall}:
\begin{quotation}
\noindent Perhaps now it is maybe expected that I should define
extension as substance or accident or else nothing at all. But by
no means, for it has its own manner of existence which fits
neither substance nor accidents [\ldots] Moreover the immobility
of space will be best exemplified by duration. For just as the
parts of duration derive their individuality from their order, so
that (for example) if yesterday could change places with today and
become the latter of the two, it would lose its individuality and
would no longer be yesterday, but today; so the parts of space
derive their character from their positions, so that if any two
could change their positions, they would change their character at
the same time and each would be converted numerically into the
other \emph{qua} individuals.  The parts of duration and space are
only understood to be the same as they really are because of their
mutual order and positions (\emph{propter solum ordinem et
positiones inter se}); nor do they have any other principle of
individuation besides this order and position which consequently
cannot be altered.
\end{quotation}

\noindent Earlier \citep{Torretti00},
Torretti had downplayed the issue of the physical individuation of
space-time points, noticing that
\begin{quotation}
\noindent [\ldots] the idea that space-time
points are what they are only by virtue of the metric structure to
which they belong agrees well with the thesis, common to Leibniz
\emph{and} Newton, that ``it is only by their mutual order and
position that the parts of time and space are understood to be the
very same which in truth they are,'' for ``they do not possess any
principle of individuation apart from this order and these
positions.''
\end{quotation}

\noindent Torretti goes on to point out that making this assumption entails
very important consequences: for instance, ``it is obviously
meaningless to speak in General Relativity of a space-time point
at which the metric is not defined,'' it becomes impossible to
hold that ``the metric of a relativistic space-time is not a
matter of fact, but of mere convention'' (geometric
conventionalism), and serious problems arise for the ``fashionable semantic theory [Kripke's] that conceives of proper names as
`rigid designators,' denoting the same individual in many
alternative diversely structured `possible worlds.' Proper names
cannot function in this way if the very individuals which are
their referents owe their identity to the structure in which they
are enmeshed.''

In conclusion, Torretti proposes a more equitable
``way of dealing with Einstein's [Hole Argument], which does not
assume that space-time points can only be physically distinguished
by means of their metric properties and relations.'' To reject the
Hole Argument, he argues, it is enough to note that two
physical objects can be distinguished either \emph{empirically}
(basically, because our direct experience suggest they differ) or
\emph{rationally} (``if they are equated to or represented by
structurally unequal conceptual systems''). The two physical
situations envisaged in the Hole Argument are both
observationally indistinguishable (in short, because of the
point-coincidence argument) and conceptually indistinguishable
(because structurally isomorphic): they are
\begin{quotation}
\noindent [\ldots] as far as our assumptions go, perfectly
indiscernible, and therefore must be regarded as identical. In the
view I have just put forward, \emph{the onus of individuating the
points of space-time does not rest with the metric, which is a
structural feature of the world. The role of structure is not to
individuate, but to specify; and of course it cannot perform this
role beyond what its own specific identity will permit, that is,
``up to isomorphism.''} It is only on nonconceptual grounds that
two isomorphic structures can be held to represent two really
different things.
\end{quotation}
In essence, in 1983 Torretti was satisfied with a
structuralist view \emph{\`a la Newton}, conjoined with the modern
understanding of \emph{Leibniz equivalence}.

However, as Friedman has remarked \citeyearpar[p.\
663]{Friedman02}, if we stick to simple Leibniz equivalence, ``how
do we describe this physical situation \emph{intrinsically}?''
What is the meaning of point-events as the local elements of
space-time?  We believe that the task of describing the physical
situation \emph{intrinsically} is worth pursuing. To this end, we
can take advantage of the fact that the points of
general-relativistic space-times, quite unlike the points of the
homogeneous Newtonian space, are endowed with a remarkably rich
\emph{non--point-like} texture\footnote{More important, as we
shall see, the \emph{physical} individuation of points as
\emph{point-events} is necessarily \emph{nonlocal} in terms of the
manifold points.} provided by the \emph{metric field}. This
texture can be exploited for the purpose of the physical
individuation of points, for it is now the \emph{dynamical} metric
field that characterizes their ``mutual order and positions.''
Furthermore, as we shall see, the need to connect the formal
structure of the theory to the empirical requirements of
measurements leads necessarily to a refinement of Leibniz
equivalence.

Following this line of thought, we shall argue that there is a
specific technical sense in which a procedure of point
individuation follows directly from the Hamiltonian formulation of
general relativity as a gauge theory. In particular, we will show
that the individuation of points originates directly from the
\emph{effective degrees of freedom of the gravitational field},
which come to play the role of \emph{basic metrical fingerprints}.

\subsection{What is the metrical fingerprint of point-events?}

Now, how is it that the metric field can realize concretely its
would-be role of physical individuator? After all, we know very well
that only a subset of the ten components of the metric is physically
significant. It seems then plausible that only this part of the metric
might serve as individuating field, while the remaining components
would carry physically spurious information.

We move from the analysis given by Bergmann and Komar,\footnote{See
\citet{Bergmann00,Bergmann01,Bergmann02,Bergmann03,
Komar00,Komar01,Bergmannkomar00,Bergmannkomar01}.} who suggest that
(in the absence of matter fields) the value of four invariant scalar
fields built from contractions of the Weyl tensor can be used as
\emph{intrinsic pseudo-coordinates} that are invariant under
diffeomorphic transformations.  Stachel \citeyearpar{Stachel02}
reprises this suggestion, but he does not pursue it further.\footnote{
To our knowledge, Bergmann and Komar did not follow up on their
suggestion, either. The last organic presentation of the issue seems to be
Bergmann's \emph{Handbuch} article \citep[p.\ 252--255]{Bergmann02}.}

Our considerations are based on the technical premises laid down by
Lusanna and Pauri \citeyearpar{Lusannapauri00} with the purpose of
extending and clarifying the Bergmann--Komar--Stachel program within
the Hamiltonian formulation of general relativity as a gauge
theory. Three circumstances make the recourse to the Hamiltonian
formalism especially propitious.
\begin{enumerate}
\item It is evident that the Hole Argument is inextricably
entangled with the initial-value problem of general relativity,
but, strangely enough, the Hole Argument has never been explicitly
discussed in that context in a systematic way. Possibly the reason
is that most authors have implicitly adopted the Lagrangian
approach (the \emph{manifold way}), where the initial-value
problem is intractable because of the nonhyperbolic nature of
Einstein's equations.\footnote{Actually, David Hilbert was the
first person to discuss the Cauchy problem for the Einstein
equations and to realize its connection to the Hole
phenomenology \citep{Hilbert00}. He discussed the issue in the
context of a general-relativistic generalization of Mie's special-relativistic nonlinear electrodynamics, and pointed out the
necessity of fixing a specific geometrically adapted
(\emph{Gaussian} in his terms, or geodesic normal as known today)
coordinate system to assure the causality of the theory. In this
connection see \cite{Howardnorton00}. However, as noted by
Stachel \citeyearpar{Stachel05}, Hilbert's analysis was incomplete and
neglected important related problems.}
\item Only in the Hamiltonian approach can we isolate the gauge
variables, which carry the descriptive arbitrariness of the
theory, from the (Dirac) observables, which are the right
candidates to become the dynamical individuating fields.
\item Finally, in the context of the Hamiltonian formalism, we can
resort to Bergmann and Komar's theory of general coordinate-group
symmetries \citep{Bergmannkomar01} to clarify the significance of
active diffeomorphisms as \emph{on-shell dynamical symmetries} of the
Einstein equations. This step is crucial: to understand fully the role
played by \emph{active} diffeomorphisms in the Hole Argument, it is
necessary to interpret them as the \emph{manifold-way} counterparts of
suitable Hamiltonian gauge transformations, which are
\emph{passive}\footnote{This passive view of active diffeomorphisms is
not equivalent to their \emph{dual} representation by the
corresponding passive coordinate transformations [as shown, for
instance, by \citet{Wald00}; see also footnote \ref{note:trick}]. By
rights, the active diffeomorphisms should be considered as passive transformations on the
function space of metric fields, rather than on the space-time
manifold.} by definition.
\end{enumerate}

\section{Mathematical development: general relativity as a gauge
theory and the physical individuation of point-events}

This section provides the technical foundations for our analysis
of the physical individuation of point-events in general
relativity. We start off with a brief, qualitative outline of
general relativity as a constrained Hamiltonian theory (especially
for the benefit of the philosophers of science who have not had
the chance of studying it in detail): Sec.\ \ref{sec:congen}
introduces constrained Hamiltonian theories in general, while
Sec.\ \ref{sec:gencon} specializes to the case of gravity. Sec.\
\ref{sec:groups} discusses the relation between the gauge
transformations of the Hamiltonian formalism and the dynamical
symmetries of the Einstein equations. Finally, in Sec.\
\ref{sec:physinv} we present the theory of the Bergmann--Komar
intrinsic coordinates, and we explore their link with gauge
freedom in general relativity and their significance for the
physical individuation of space-time points.

\subsection{The constrained Hamiltonian formalism}
\label{sec:congen}

As most other fundamental theories in modern physics, general
relativity falls under the chapter of \emph{gauge theories}. To
use the very general definition given by Henneaux and Teitelboim
\citeyearpar{Henneauxteitelboim00}:

\begin{quotation}
\noindent These are theories in which the physical system being
dealt with is described by more variables than there are
physically independent degrees of freedom. The physically
meaningful degrees of freedom then reemerge as being those
invariant under a transformation connecting the variables (gauge
transformation). Thus, one introduces extra variables to make the
description more transparent, and brings in at the same time a
gauge symmetry to extract the physically relevant content.
\end{quotation}

\noindent The mathematical development of gauge theories starts when
we realize that the Lagrangian action principle, $\delta \int
\mathcal{L}(q,\dot{q}) \, dt = 0$, yields Euler--Lagrange equations
that are not hyperbolic, because they cannot be solved for all the
accelerations. Technically, the same condition that makes it so (the
singularity of the Hessian matrix\footnote{Throughout this section we
shall outline the constrained Hamiltonian theory in the simpler case
of a finite number of degrees of freedom. For field theories (such as
general relativity) there are, as always, additional subtleties.}
$\partial^2 \mathcal{L}/\partial \dot{q}^k \partial \dot{q}^h$) means
also that, when we move from the Lagrangian to the Hamiltonian
formulation, the momenta are not all functionally independent, but
satisfy some conditions known as (\emph{primary}) \emph{constraints}.
\emph{Secondary} constraints arise when we require that the primary
constraints be preserved through evolution.\footnote{Tertiary
constraints follow from the conservation of secondary constraints, and
so on. In physically interesting theories this chain ends before we
run out of all the original degrees of freedom.} There is no strong
distinction between primary and secondary constraints in the role that
they come to play in the unfolding of constrained dynamics.

The existence of constraints implies that not all the points of phase
space represent physically meaningful states: rather, we are
restricted to the \emph{constraint surface} where all the constraints
are satisfied.  The dimensionality of the constraint surface is given
by the number of the original canonical variables, minus the number of
functionally independent constraints.

Generally, the constraints are given as functions of the canonical
variables which vanish on the constraint surface; technically, these
functions are said to be \emph{weakly} zero\footnote{Conversely, any
\emph{weakly} vanishing function is a linear combination of the
\emph{weakly} vanishing functions that define the constraint surface.}
($\approx 0$).  Note that weakly vanishing functions may have
nonvanishing derivatives in directions normal to the constraint
surface, so their Poisson brackets with some of the canonical
variables may well be nonzero.  If instead all the derivatives vanish,
the functions are said to be \emph{strongly} zero, and they can be freely
inserted in any Poisson bracket without changing the result.

When used as generators of canonical transformations, some
constraints, known as \emph{first class},\footnote{A function of the
canonical variables is defined to be \emph{first class} if its Poisson
brackets with all the constraints are strongly or weakly zero. It is defined to be
\emph{second class} if its Poisson bracket with at least one
constraint is not zero.} will map points on the constraint surface to
points on the same surface; these transformations are known as
\emph{gauge transformations}. \emph{Second class} constraints, on the
contrary, will generate transformations that map allowed physical
states (points on the constraint surface) onto disallowed states
(points off the constraint surface). Since second-class constraints do
not show up in the Hamiltonian formulation of general relativity, we
will disregard them in the rest of this exposition.

To obtain the correct dynamics for the constrained system, we need
to modify the Hamiltonian variational principle to enforce the
constraints; we do this by adding the constraint functions to the
Hamiltonian, after multiplying them by arbitrary functions of time
(the \emph{Lagrange--Dirac multipliers}). Because the first-class
constraints generate gauge transformations on the constraint
surface, different choices of the Lagrange--Dirac multipliers will
generate evolutions of the canonical variables that differ by
gauge transformations. If, with Dirac, we make the reasonable demand
that the evolution of all \emph{physical variables} should be
unique,\footnote{Otherwise we would have to envisage \emph{real}
physical variables that are indeterminate and therefore not
\emph{observable}, and ultimately not measurable.} then the points
of the constraint surface that sit on the same \emph{gauge orbit}
(that is, that are linked by gauge transformations) \emph{must
describe the same physical state}. Conversely, only the functions
in phase space that are invariant with respect to gauge
transformations can describe physical quantities.\footnote{Of
course, in many cases (such as electromagnetism) we know the
observable quantities from the beginning, because we have
gauge-independent dynamical equations for the fields (e.\ g., the
Maxwell equations). Then the distinction between observables and
gauge variables that follows from the first-class constraints must
reproduce this situation.}

To eliminate this ambiguity and create a one-to-one mapping
between points in phase space and physical states, we can impose
further constraints, known as \emph{gauge conditions}. The gauge
conditions can be defined by arbitrary functions of the variables of the
constraint surface, except that they must define a \emph{reduced
phase space} that intersects each gauge orbit exactly once. In
other words, given a point on the constraint surface, there must
be a gauge transformation that takes it into the reduced phase
space; conversely, if we apply a gauge transformation to a point
in the reduced phase space, we take it out of the gauge.
Abstractly, reduced phase space is the quotient of the constraint
surface by the group of gauge transformations and it represents
the space of variation of the true degrees of freedom of the theory.

The number of independent gauge conditions must be equal to the number
of independent first-class constraints.  Because of their role, the
gauge conditions cannot commute (have vanishing Poisson bracket) with
the original first-class constraints; so the set of the first-class
constraints, with the addition of the gauge conditions, becomes a set
of \emph{second-class} constraints.  After this \emph{canonical
reduction} is performed, the theory is completely determined: each
physical state corresponds to one and only one set of canonical
variables that satisfy the constraints and the gauge conditions. Then we
are also able to determine the Lagrange--Dirac multipliers, so no
arbitrary functions of time appear anymore in the Hamiltonian.

At this stage, we can invoke the \emph{Shanmugadhasan transformation}
\citep{Shanmugadhasan00} to put the gauge conditions into an
especially meaningful functional form. The Shanmugadhasan
transformation has the effect of reshuffling all the first-class
constraints into a set of Abelian canonical momenta. The surface where
these momenta are zero is just the original constraint surface, and
the conjugate canonical variables are the gauge functions, whose gauge
fixing determines the reduced phase space. The so-called \emph{Dirac
observables} are just a Darboux basis for the reduced phase
space.\footnote{While the Poisson brackets of the \emph{Dirac}
observables with the original constraints vanished only \emph{weakly},
the reduced phase space is equipped with a new Poisson--Dirac algebra
given by the so-called \emph{Dirac brackets} (denoted by
$\{\cdot,\cdot\}^*$), and the Dirac brackets of the observables with
the \emph{Abelianized} constraints and their conjugate variables
vanish \emph{strongly}. This is precisely the purpose of the
Shanmugadhasan transformation, which creates a true \emph{projection}
from the original constraint surface to the reduced phase space.} Note
that the entire procedure of canonical reduction is performed
\emph{off shell}, that is, without reference to the actual solution of
the Hamilton equations.

Thus, after reducing twice the dimension of the initial phase
space by the number of independent constraints (once to go to the
constraint surface, once again when the gauge conditions are
enforced to obtain the reduced phase space), we are at the end of
our long trip. Under the action of the Hamiltonian, the Dirac
observables \emph{evolve deterministically} within the reduced
phase space, and the indeterminateness of the nonhyperbolic
Euler--Lagrange equations has been converted into the physically
harmless arbitrariness of the gauge fixing.

\subsection{General relativity as a constrained Hamiltonian theory}
\label{sec:gencon}

The standard progression of general-relativity textbooks takes us
through a dense barrage of differential geometry until we have
gathered enough foundations to lay down the vacuum Einstein equations,
\begin{equation}
R_{\mu \nu} - \frac{1}{2} R g_{\mu \nu} = 0; \label{eq:einstein}
\end{equation}
on this mountaintop we can draw a breath of relief, and
contemplate the beauty of general relativity. These equations can
be derived as Euler--Lagrange equations from the Lagrangian
variation of the Einstein--Hilbert action
\begin{equation}
S = \int d^4x \, \sqrt{-g} R,
\end{equation}
where the independent components of the metric field $g_{\mu \nu}$
serve as configuration variables. However, the Eqs.\
\eqref{eq:einstein} cannot be solved as they are written, because
they are not hyperbolic: only two equations out of ten are
evolution equations for the ``accelerations'' of the metric.  The
reason is that the action is invariant under general coordinate
transformations (the \emph{passive diffeomorphisms} $\passdiff$),
so the Hessian matrix has vanishing determinant
\citep{Sundermayer00}. From the Lagrangian point of view, to solve
the Eqs.\ \eqref{eq:einstein} we need to remove the diffeomorphism
invariance by fixing the coordinate system completely.\footnote{In
the Lagrangian formalism (\emph{manifold way}), the counting of
degrees of freedom goes as follows: the ten Einstein equations can
be rearranged as four \emph{Lagrangian constraints} (restrictions
on the initial Cauchy data), four \emph{Bianchi identities} (which
vanish identically), and two dynamical second-order equations.
Therefore, of the ten independent components of the metric tensor,
two are deterministic dynamical degrees of freedom, four are bound
by the Lagrangian constraints, and the remaining four are
completely indeterminate until the coordinates are fixed.}

Let us now turn to the Hamiltonian formalism, where the gauge
symmetry of the system is fully manifest. Although several
variations are possible, we will outline the standard ADM
formalism [named after Arnowitt, Deser and Misner \citeyearpar{Adm00}].
Before we attempt to solve the Cauchy problem for the Einstein
equations, we need to perform a $3+1$ \emph{split} of space-time:
that is, we need to assume that the space-time $(\mfour,{}^4 g)$
is globally hyperbolic, and that it can be \emph{foliated} by a family
of spacelike Cauchy surfaces
$\Sigma_{\tau}$, indexed by the \emph{parameter time} $\tau$. This
means essentially that we view the global space-time as
representing the (parameter) time development of a
three-dimensional Riemannian metric ${}^3 g$ on a fixed
tridimensional manifold $\Sigma_{\tau}$. The three-metric is a
classical field which depends on the three spatial
coordinates\footnote{From now on, it will be our convention to
drop all the ``${}^3$'' indices which denote tensors on the
spatial manifold; furthermore, we will use lowercase Latin indices
to enumerate the spatial coordinates, and uppercase Latin indices
to enumerate parameter time plus the spatial coordinates.}
$\sigma^a$ on $\Sigma_{\tau}$, and evolves with the parameter time
$\tau$.

To complete the $3+1$ split, we need to specify the packing of the
surfaces $\Sigma_\tau$ in proper (physical) time, and the
\emph{physical correspondence} between the points on each surface
(loosely, we need to keep track of which point is which as we
progress through time). These choices are achieved by specifying
the \emph{lapse function} $N$ and the shift vector $N^a$. Only now
the four-metric can be reconstructed from the $\tau$ dependence of
the three-metric, the lapse, and the shift.

The $(3+1)$-split Einstein equations are obtained from the
Lagrangian variation of the \emph{ADM action},
\begin{equation}
S_{ADM} = \int d\tau \, N \int_{\Sigma_\tau} d\sigma^a \sqrt{-g} \,
[R + K_{ab} K^{ab} - K^2] + \mathrm{surface \,
terms}\footnote{Within the rest of this paper, we shall always
neglect these terms.},
\end{equation}
where $R$ is the scalar curvature of the three-metric $g_{ab}$,
where the \emph{extrinsic curvature} $K_{ab}$ is essentially the
$\tau$ derivative of $g_{ab}$, and where $K = {K_a}^a$. The ten
configuration variables are $N$, $N^a$, and the six independent
components of $g_{ab}$. The Legendre transformation yields the
momenta
\begin{equation}
\pi^{ab} = -\sqrt{-g} [K^{ab} - K g^{ab}] \quad
(\mathrm{conjugated\ to\ } g_{ab}),
\end{equation}
\begin{equation}
\label{eq:pizero}
\pi_0 = 0, \; \pi_a = 0 \quad (\mathrm{conjugated\ to\ } N, N^a).
\end{equation}
Phase space is indexed by the 20 variables $(N,\pi_0)$,
$(N^a,\pi_a)$, $(g_{ab},\pi^{ab})$, but the conditions \eqref{eq:pizero} on the
momenta conjugated to lapse and shift must be understood as the
\emph{primary constraints} of the theory, and therefore should be
written as $\pi^A \approx 0$. By requesting that the primary
constraints be preserved through dynamical evolution, we obtain
the secondary constraints,
\begin{equation}
\mathcal{H}_0 \equiv \frac{1}{\sqrt{-g}} \left[ \pi^{ab} \pi_{ab} -
\frac{1}{2} ({\pi_a}^a)^2 \right] - \sqrt{-g} \, R \approx 0 \quad
(\mathrm{superhamiltonian\ constraint}),
\end{equation}
\begin{equation}
\mathcal{H}_a \equiv - 2 {{\pi_a}^b}_{|b} \approx 0 \quad
(\mathrm{supermomentum\ constraints}),
\end{equation}
where the bar denotes covariant differentiation on $\Sigma_\tau$.
Altogether, the primary and secondary constraints restrict the
allowable physical states to a 12-dimensional constraint surface
$\Gamma_{12}$ in phase space. The $\pi_A$ and the $\mathcal{H}_A$
are all first-class constraints, and generate \emph{gauge}
transformations on the constraint surface: the effect of the
$\pi_A$ is to change the lapse and shift, while $\mathcal{H}_0$
and the $\mathcal{H}_a$ respectively induce normal deformations of
the surfaces $\Sigma_\tau$, and generate transitions from a
three-coordinate system to another. There are no second-class
constraints.

The \emph{Dirac Hamiltonian} (which rules the constrained
dynamics) can be written purely in terms of the
constraints:\footnote{Even before adding the constraints, the
canonical Hamiltonian can be written as $H_C = \int d\sigma^a N^A
\mathcal{H}_A$, so we could formally absorb the Lagrange--Dirac
multipliers relative to the $\mathcal{H}_A$ into the definition of
the $N^A$. Still, lapse and shift are not arbitrary functions,
but dynamical variables! The fact that the Hamiltonian
vanishes on the constraint surface is a general feature of
generally covariant theories. See for instance
\citet{Henneauxteitelboim00}.}
\begin{equation}
H_D = \int d\sigma^a [ N^A \mathcal{H}_A + \lambda^A \pi_A ],
\end{equation}
where the $\lambda^A$ are Lagrange--Dirac multipliers. At this
stage we have already restored the hyperbolicity of the (Hamilton)
equations of motion, but at the price of introducing
the four arbitrary functions of time\footnote{The $\lambda^A$ are
also arbitrary functions of the spatial coordinates $\sigma^a$,
although in a slightly different sense: loosely speaking, there
are four arbitrary multipliers \emph{at each spatial location},
so the spatial coordinates, together with ``${}^A$'', play the
role of generalized degree-of-freedom indexes.} $\lambda^A$:
\begin{equation}
\dot{N}^A \approx \lambda^A, \quad \dot{g}_{ab} \approx
f_{ab}[g,\pi|\lambda],
\end{equation}
\begin{equation}
\dot{\pi}^A \approx 0, \quad \dot{\pi}_{ab} \approx
h_{ab}[g,\pi|\lambda].
\label{eq:afterthis}
\end{equation}
To remove this arbitrariness, we must fix the gauge as follows. The
first step is the gauge fixing to the secondary constraints: we choose
four functions $\chi_A$ of the $g$ and $\pi$ (but not of $N^A$!) that
satisfy the orbit conditions,\footnote{These conditions implement the
Lorentz signature of the reconstructed four-metric, by inheriting the
signature already implicit in the superhamiltonian and supermomentum.}
$\mathrm{det} \, |\{\chi_A,\mathcal{H}_B\}| \neq 0$, and we impose
$\chi_A \approx 0$ on the constraint surface. It turns out that the
requirement of time constancy for the gauge fixings $\chi_A$ fixes the
gauge with respect to the primary constraints.  Finally, the
requirement of time constancy for these latter gauge fixings
determines the multipliers $\lambda^A$. So the choice of the four
constraints $\chi_A$ is sufficient to remove all the gauge
arbitrariness.

Under the Shanmugadhasan transformation proposed by Lusanna \citeyearpar{Lusanna00,Lusanna01}, the superhamiltonian constraint corresponds\footnote{In practice, this transformation requires the solution of the superhamiltonian constraint, but so far this result has proved elusive.} to a new canonical pair: the unknown variable in which the constraint must be solved is the conformal factor of $g$ (proportional to $\mathrm{det} \, g$), while the gauge parameter is the conformal-factor momentum $\pi_\phi$ (which determines the normal deformations of $\Sigma_\tau$). The corresponding gauge fixing, $\chi_0 \approx 0$, has the effect of selecting the shape of $\Sigma_\tau$. The supermomentum constraints correspond to three canonical pairs, namely the three longitudinal components of $\pi^{ab}$, and three gauge parameters, namely the three-coordinates on $\Sigma_\tau$. The corresponding gauge fixings, $\chi_a \approx 0$, have the effect of selecting the coordinate system on $\Sigma_\tau$. After the gauge parameters have been fixed, the second-order time-constancy requirement (mentioned above) has the effect of providing partial differential equations for the lapse and shift, in a manner compatible with the shape of $\Sigma_\tau$ and with the choice of the three-coordinates.

At the end of the canonical reduction procedure, the 12 degrees of
freedom of the constraint surface are reduced to four, the
\emph{Dirac observables} $q^r, p_s$ ($r,s=1,2$) that index the
reduced phase space $\Psi_4$, and that represent the two
\emph{true dynamical degrees of freedom of the gravitational
field}. Each gauge fixing creates a realization of
$\Psi_4$, with a canonical structure implemented by the Dirac
brackets associated to that gauge. The Dirac observables
satisfy the final Hamilton equations,
\begin{equation}
\dot{q}^r = \{q^r, E_{\mathrm{ADM}}\}^*, \quad \dot{p}_s = \{p_s,
E_{\mathrm{ADM}}\}^*,
\end{equation}
where $E_{\mathrm{ADM}}$ is intended as the restriction of the ADM
Energy to $\Psi_4$ and where the $\{\cdot,\cdot\}^*$ are the Dirac
Brackets. In general, $q^r(\tau,\sigma^a)$ and
$p_s(\tau,\sigma^a)$ are highly nonlocal\footnote{Because in
general relativity the Shanmugadhasan transformation is highly
nonlocal.}; \emph{a priori} they are neither tensors nor invariants under
space-time diffeomorphisms, because their functional form depends
on the gauge fixing. As we shall see, \emph{on shell} (when the
dynamical variables are restricted to the values that they can
have as solutions of the Hamilton--Dirac equations) the gauge
fixing is equivalent to the choice of a set of four-dimensional
coordinates.

According to Lusanna and Pauri \citeyearpar{Lusannapauri00}, the
Shanmugadhasan transformation proposed by
Lusanna \citeyearpar{Lusanna00,Lusanna01} allows the (loose)
interpretation of the Dirac observables as representing the
\emph{tidal effects} of the gravitational field. Obviously, in
general relativity there are no gravitational \emph{forces} in the common sense.
Yet, we can introduce the general-relativistic analogs of inertial
forces with respect to the worldlines of nongeodesic observers
\citep{Abramowicz00,Abramowicznurowskiwex00}. The physical meaning
of the eight gauge transformations is just to modify the
\emph{inertial} (reference-frame--induced) effects; however, the
\emph{presentation} of both the tidal effects and the inertial
forces depends on the gauge fixings, just as the functional form
of the Dirac observables does.

\subsection{Gauge groups and dynamical symmetries in the general theory of relativity}
\label{sec:groups}

Not all the transformations generated by the first-class constraints
(the \emph{off-shell Hamiltonian gauge group} $\mathcal{G}_8$) are
true, \emph{harmless} gauge transformations in the sense introduced by Dirac,
because some of them will join points of the constraint surface
that represent different four-geometries,\footnote{The quotient of
the constraint surface with respect to the off-shell Hamiltonian
gauge transformations is the so-called \emph{reduced off-shell
conformal superspace} $\Gamma_4 = \Gamma_{12}/\mathcal{G}_8$. Each
point of $\Gamma_4$ (a \emph{Hamiltonian off-shell} or
\emph{kinematical} gravitational field) is an equivalence class known
as \emph{off-shell conformal three-geometry} for the space-like
hypersurfaces $\Sigma_\tau$. It is not a four-geometry, because it
contains all the off-shell three-geometries connected by
Hamiltonian gauge transformations.} and therefore different
physical states.  This property follows from the fact that, in the
Dirac Hamiltonian, among the eight multipliers only four are
arbitrary Lagrange--Dirac multipliers (the other four are the
dynamical variables lapse and shift), and that the correct
gauge-fixing procedure starts by giving only the four gauge
fixings for the secondary constraints. Going \emph{on shell}
(that is, restricting our consideration to the solutions of
the Hamilton--Dirac equations) we introduce a functional dependence
among the group descriptors of $\mathcal{G}_8$, creating a
four-dimensional subgroup $\mathcal{G}_4^\mathrm{dyn}$ (the
\emph{on-shell Hamiltonian gauge group}) whose transformations are
also \emph{dynamical symmetries} of the Hamilton--Dirac equations (dynamical symmetries are defined as the transformations
that map solutions of the equations of motion onto other
solutions; as such, they are an on-shell concept).

In the context of the \emph{Lagrangian} formalism, the (passive)
dynamical symmetries of the Einstein equations were studied by
Bergmann and Komar \citeyearpar{Bergmannkomar01}, who showed that the
largest group of such transformations is not $\passdiff$ [${\xi'}^\mu = f^\mu(\xi^\nu)$] but rather the group $Q$ of
transformations of the form ${\xi'}^\mu = f^\mu(\xi^\nu,g_{\alpha
\beta})$. These transformations map points on points, but
associate with a given point $x$ an image point $x'$ that
depends also on the metric field $g$. Hence the elements of $Q$ should
be considered as mappings from the functional space of metric
fields onto itself.

Bergmann and Komar showed that the passive diffeomorphisms,
$\passdiff$, are a \emph{nonnormal subgroup} of $Q$. We have just
met another nonnormal subgroup of $Q$: it is the on-shell
Hamiltonian gauge group $\mathcal{G}_4^\mathrm{dyn}$, or rather
its \emph{Legendre pullback} to configuration space, which
Bergmann and Komar call $Q_\mathrm{can}$. The subgroups
$\passdiff$ and $Q_\mathrm{can}$ have a nonempty intersection,
which consists of all the passive coordinate transformations that
respect the $3+1$ splitting of the ADM construction.

Looking in the other direction (from configuration space to phase
space), $Q_\mathrm{can}$ represents the part of $Q$ that is
\emph{projectable} into phase-space transformations. It follows
that the subgroup $Q_\mathrm{can}$ is defined by a particular
choice of the four functionally independent descriptors that are
the manifold counterparts of the four independent descriptors of
$\mathcal{G}_4^\mathrm{dyn}$.

All these groups are just different representations of
the descriptive arbitrariness of general relativity, so we expect
that they should all generate the same partition of the
space $\riem$ of solutions of the Einstein--ADM equations
into equivalence classes. Indeed, Bergmann and Komar showed that
\begin{equation}
\geom = \frac{\riem}{\passdiff} = \frac{\riem}{Q_\mathrm{can}} = \frac{\riem}{Q},
\label{eq:equno}
\end{equation}
which is mathematically possible because both $\passdiff$ and
$Q_\mathrm{can}$ are nonnormal subgroups of Q.

Only one detail is missing: what is the status of the active
diffeomorphisms $\actdiff$ within this representation? Intuitively, it
seems that active and passive diffeomorphisms make up all the
operations that can be defined on the space-time manifold; however,
nobody so far has studied in detail the mathematical structure of the
group $Q$. It is however easy to show \citep{Lusannapauri00} that at
least the infinitesimal active diffeomorphisms belong to $Q$, because
they can be interpreted as passive transformations with the following
procedure.

Consider an infinitesimal (passive) transformation of the type
${\xi'}^\mu = \xi^{\mu} + X^{\mu}(\xi,g)$. This will induce the usual
formal local variation of the metric tensor,
\begin{equation}
\bar{\delta} g_{\mu\nu} =
-\big( X_{\mu;\nu}(\xi,g) + X_{\nu;\mu}(\xi,g) \big).
\end{equation}
Therefore, if $\bar{\delta} g_{\mu\nu}$ is the variation of the
metric tensor associated with the infinitesimal \emph{active}
diffeomorphism, the solution $X^{\mu}(\xi,g)$ of these
Killing-type equations identifies a corresponding passive Bergmann--Komar dynamical symmetry of $Q$.
This should imply that all the active diffeomorphisms connected
with the identity in $\actdiff$ can be reinterpreted as
elements of a nonnormal subgroup of the generalized passive
transformations of $Q$. Clearly this subgroup is disjoint from the
subgroup $\passdiff$: note that this is possible because
diffeomorphism groups do not possess a canonical identity. Given
this, we could naturally guess that $Q_\mathrm{can}$ is a mix of
passive and active diffeomorphisms, because the active and passive
diffeomorphisms, being nonnormal subgroups of $Q$, should, as it
were, fill $Q$ densely in a suitable topology.

Finally, we complete Eq.\ \eqref{eq:equno}: because obviously we have
\begin{equation}
\geom = \frac{\riem}{\passdiff} = \frac{\riem}{\actdiff},
\end{equation}
we obtain the final definition of the equivalence classes
of \emph{on-shell} or \emph{dynamical} gravitational fields,
\begin{equation}
\geom = \frac{\riem}{\passdiff} = \frac{\riem}{\actdiff} = \frac{\riem}{Q_\mathrm{can}} = \frac{\riem}{Q}.
\label{eq:eqdue}
\end{equation}
In other words, any of the groups $\passdiff$, $\actdiff$,
$Q_\mathrm{can}$, and $Q$ can be used to implement Leibniz
equivalence \emph{on shell}.

\subsection{The Bergmann--Komar invariants: metrical structure
and the physical individuation of points in the (un)real world}
\label{sec:physinv}

Let us now take a quick detour back to four-dimensional (so to speak) general relativity. We note with Bergmann and Komar\footnote{See \citet{Bergmann00,Bergmann01,Bergmann02,Bergmann03,Bergmannkomar00}.} that for a vacuum solution of the Einstein equations, in the hypothesis that space-time admits \emph{no symmetries}, there are exactly four functionally independent scalars that can be written using the lowest possible derivatives of the metric.\footnote{The fact that there are just \emph{four} independent invariants is crucial for the purpose of point individuation, and it should not be regarded as a coincidence. After all, recall that in general space-times with matter there are 14 invariants of this kind! \citep{Geheniaudebever00}} These are the four Weyl scalars (the eigenvalues of the Weyl tensor), here written in Petrov compressed notation,
\begin{eqnarray}
w_1 &=& \mathrm{Tr} \, (g W g W), \\
w_2 &=& \mathrm{Tr} \, (g W \epsilon W), \\
w_3 &=& \mathrm{Tr} \, (g W g W g W), \\
w_4 &=& \mathrm{Tr} \, (g W g W \epsilon W),
\end{eqnarray}
where $g$ is the \emph{four}-metric, $W$ is the Weyl
tensor, and $\epsilon$ is the Levi--Civita totally antisymmetric
tensor.

Bergmann and Komar then propose that we build a set of \emph{intrinsic coordinates} for the point-events of space-time as
four functions of the $w_T$,
\begin{equation}
\hat{I}^{[A]} = \hat{I}^{[A]} \bigr[ w_T[g(x),\partial g(x)] \bigl].
\end{equation}
Indeed, under the hypothesis of no space-time
symmetries,\footnote{Our attempt to use intrinsic coordinates
to provide a physical individuation of point-events would
\emph{prima facie} fail in the presence of symmetries,
when the $\hat{I}^{[A]}$ become degenerate. This objection
was originally raised by Norton \citeyearpar{Norton05} as a critique
to manifold-plus-further-structure (MPFS) substantivalism
[according to which the points of the manifold, conjoined
with additional local structure such as the metric field,
can be considered physically real; see for instance \citet{Maudlin00}].
Several responses are possible. First, although to this
day all the \emph{known} exact solutions of the Einstein
equations admit one or more symmetries, these mathematical
models are very idealized and simplified; in a realistic
situation (for instance, even with two masses) space-time
is filled with the excitations of the gravitational degrees
of freedom, and admits no symmetries at all. Second, the
parameters of the symmetry transformations can be used
as supplementary individuating fields, since, as noticed by
Stachel \citeyearpar{Stachel02}, they also depend on metric field, through its isometries.
Third, and most important,
in our analysis of the physical individuation of points
we are arguing a question of principle, and therefore we
must consider \emph{generic} solutions of the Einstein
equations rather than the null-measure set of solutions
with symmetries.} the $\hat{I}^{[A]}$ can be used to label
the point-events of space-time, at least
locally.\footnote{Problems might arise if we try to extend
the labels to the entire space-time: for instance, the
coordinates might turn out to be multivalued.}
What is more, the value of the intrinsic coordinates at
a point-event can be extracted (in principle) by an actual experiment
designed to measure the $w_T$ (see Sec.\ \ref{sec:gps}).
Because they are functionals of scalars, the $\hat{I}^{[A]}$
are invariant under passive diffeomorphisms (therefore they
do not define a coordinate chart in the usual sense), and
by construction they are also constant under the drag-along
of tensor fields induced by active
diffeomorphisms.\footnote{Already at this stage, we see that
this is just the right method to realize the equivalence
class of points to which Bartels was alluding \citep{Bartels00}.}

The metric can be rewritten with respect to the intrinsic
coordinates:
\begin{equation}
\hat{g}^{[AB]} = \frac{\delta \hat{I}^{[A]}}{\delta x^\mu}
\frac{\delta \hat{I}^{[B]}}{\delta x^\nu} g^{\mu \nu}.
\label{eq:deceiving}
\end{equation}
The $\hat{g}^{[AB]}$ represent the ten \emph{invariant scalar
components} of the metric; of course they are not all independent,
but they should satisfy six functional restrictions that follow
from the Einstein equations. However, Eq.\ \eqref{eq:deceiving} is
deceiving, because the $\hat{g}^{[AB]}$ are functionals of the
metric and of \emph{its partial derivatives} (through the
$\hat{I}^{[A]}$). It should be noted that, in a sense, the freedom
to express the metric using any set of coordinates is still
present in the choice of the four functions $\hat{I}^{[A]}$ of the
Weyl scalars. What is more, given any coordinatization of a
space-time without symmetries, it is possible to reproduce the
tensorial components of the metric using a suitable set of
$\hat{I}^{[A]}$.

Decomposing the $w_T$ with the $3+1$ splitting outlined in Sec.\
\ref{sec:gencon}, we realize [again with Bergmann and Komar
\citeyearpar{Bergmannkomar00}] that the four Weyl scalars $w_T$
\emph{do not depend on lapse and shift}. This circumstance is crucial,
because it means that \emph{we can use suitable functions of the $w_T$
as gauge fixings to the secondary constraints}\footnote{Please refer
back to Sec.\ \ref{sec:gencon}, just after Eq.\ \eqref{eq:afterthis}.}
\citep{Lusannapauri00}.  To do so, we first write the
Bergmann--Komar intrinsic coordinates as functionals of the ADM
variables,
\begin{equation}
\hat{I}^{[A]} [w_T(g,\partial g)] \equiv \hat{Z}^{[A]}
[w_T(g,\pi)];
\end{equation}
we then select a \emph{completely arbitrary} coordinate system
$\sigma^A \equiv [\tau,\sigma^a]$ adapted to the $\Sigma_\tau$
surfaces; finally, we apply the gauge fixing $\Gamma$ defined by
\begin{equation}
\chi^A \equiv \sigma^A - \hat{Z}^{[A]}
\bigl[ w_T[(g(\sigma^B),\pi(\sigma^C)] \bigr] \approx 0;
\end{equation}
of course the functions $\hat{Z}^{[A]}$ must be chosen to satisfy
the orbit conditions $\{\hat{Z}^{[A]},\mathcal{H}_B\} \neq 0$,
which ensure the independence of the $\chi^A$ and carry
information about the Lorentz signature.  The effect is that the
evolution of the \emph{Dirac observables}, whose dependence on
space (and on parameter time) is indexed by the chosen coordinates
$\sigma^A$, reproduces the $\sigma^A$ as the Bergmann--Komar
intrinsic coordinates:
\begin{equation}
\sigma^A = \hat{Z}^{[A]} [w_T(q^r(\sigma^B),p_s(\sigma^C)|\Gamma)],
\label{eq:superdef}
\end{equation}
where the notation $w_T(q,p|\Gamma)$ represents the functional form
that the Weyl scalars $w_T$ and the \emph{Dirac observables} $q^r$,
$p_s$ assume in the chosen gauge. Eq.\ \eqref{eq:superdef} is just an
identity with respect to the $\sigma^A$. The price that we have paid
for this achievement is of course that we have broken general
covariance!

At first this result may sound surprising:
diffeomorphism-invariant quantities, such as the intrinsic
coordinates, are known as \emph{Bergmann observables}, and are
often identified with the only locally measurable variables of the
pure gravitational field (because being diffeomorphism invariants
they can be obtained using the coordinate system corresponding to
any experimental arrangement).  From the Hamiltonian viewpoint,
however, they are gauge-dependent\footnote{\emph{Canonical
reduction} (which creates the distinction between gauge-dependent
quantities and Dirac observables) is made \emph{off shell}, that
is, \emph{before} solving the equations of motion. It is not known
so far whether suitable diffeomorphism-invariant intrinsic
coordinates can also become \emph{Dirac observables} \emph{on
shell}, that is, on the space of solutions to the equations of
motion. See however Sec.\ \ref{sec:conclusion}.} quantities that (in
a sense) can be arranged to assume any functional dependence on
$\Sigma_\tau$.

The crucial point to remember here is that the gauge
transformations of $\mathcal{G}_8$ can actually link different
four-geometries; correspondingly, a complete gauge fixing can
modify the value of diffeomorphism-invariant
quantities.\footnote{Each three-metric in the conformal gauge orbit
has a different three-Riemann tensor, and different three-curvature
scalars. Since four-tensors and four-curvature scalars depend on lapse,
shift, their gradients, and on the conformal-factor momentum, most
of these objects are in general gauge variables from the
Hamiltonian point of view.} So we can take any four-geometry, find
its Cauchy data on $\Sigma_\tau$, and then move along its
$\mathcal{G}_8$ gauge orbit to create any arbitrary structure for
the Weyl scalars; but the final point on the constraint surface
will represent a \emph{different} four-geometry. On the other
hand, the on-shell Hamiltonian gauge group
$\mathcal{G}_4^\mathrm{dyn}$ contains only transformations that
are counterparts of active or passive \emph{projectable}
diffeomorphisms (the ones that are compatible with the $3+1$ split).

After canonical reduction and only for the solutions of the equations
of motion, Eq.\ \eqref{eq:superdef} becomes a \emph{strong}
relation, and it amounts to a \emph{definition of the four coordinates
$\sigma^A$}, providing a \emph{physical individuation of any
point-event, in the gauge-fixed coordinate system, in terms of
the true dynamical gravitational degrees of freedom}.

The virtue of this elaborate setup is not that it selects a set of
physically preferred coordinates, because by modifying the
functions $I^{[A]}$ we have the possibility of implementing any
coordinate transformation. So diffeomorphism invariance reappears
under a different semblance: we find exactly the same functional
freedom whether we choose a set of coordinates on $\mfour$, the
functions $Z^{[A]}$, or the gauge fixing. Thus, it turns out that,
\emph{on shell}, at the Hamiltonian level as well as the
Lagrangian level, \emph{gauge fixing is clearly synonymous with
the selection of manifold coordinates}.
Instead, we are now able to claim that \emph{any} coordinatization
of the manifold can be seen as embodying the physical
individuation of points, because it can be
implemented\footnote{Again, at least locally.} as the
Komar--Bergmann intrinsic coordinates after we choose the correct
$Z^{[A]}$ and we select the correct gauge. The byproduct of the
gauge fixing is the identification of the form of the physical
degrees of freedom as nonlocal functionals of the metric and
curvature.

Summarizing, each of the point-events of space-time is endowed with
its own physical individuation (the right metrical fingerprint!)  as
the value, as it were, of the four canonical coordinates (just four!),
or \emph{Dirac observables} which describe the dynamical degrees of
freedom of the gravitational field. However, these degrees of freedom
are unresolveably entangled with the structure of the metric manifold
in a way that is strongly gauge dependent.

As a final consideration, let us point out that Eq.\
\eqref{eq:superdef} is a numerical identity that has an \emph{inbuilt
noncommutative structure}, deriving from the Dirac--Poisson structure
on its right-hand side. The meaning of this structure is not clear at
the classical level, but we believe that it could be relevant to the
quantization of general relativity.

\section{The individuation of points in the real world}
\label{sec:gps}

The philosophical analysis of the general-relativistic notion of
space-time is developed most often (and this paper is no exception) on
the \emph{geometrodynamical} formulation of general relativity, which
pictures matter following the straightest lines, so to speak, in a
curved space-time arena deformed by gravitation. There are many
reasons for this preference: the geometric theory is indeed very
beautiful, and it appears to complete and extend more fully the
critique of space-time structure begun with special relativity.
Within this paradigm, the prototype solution is a strongly curved
vacuum space-time with no symmetries. For such a space-time,
coordinate system are freely interchangeable, and of course they are
almost completely irrelevant to the physical individuation of
points. For such a space-time, the philosophical arguments about the
Hole Argument and about general covariance carry their full weight.

However, our universe is not a strongly curved space-time, and it is
not a vacuum solution: rather, it resembles most closely the flat
space-time of special relativity, and it contains much matter,
organized in structures at many scales. Although we know, in theory,
that all coordinate frames are equally acceptable, in this real
physical world we manage to keep the time, keep our orientation,
navigate the solar system, and make sense of the universe with a
handful of very special coordinate systems.  These systems are
precisely the ones that recognize that gravity is weak (so it can be
treated as a correction to flat space-time) and that matter with
structure is available to provide useful points of reference (in a relational
sense).

Indeed, Soffel \citeyearpar{Soffel00} defines the purpose of
\emph{astrometry} (the theory of constructing reference frames) as
``the materialization of a global, nonrotating, quasi-inertial
reference frame, in the form of a fundamental catalogue of stellar
positions and proper motions.'' On a smaller scale, the preferred
reference frames are those that provide a simple, understandable form
for the dynamical equations that rule the motions of celestial
bodies. In the case of the solar system, a suitable reference frame is
the barycentric post--Newtonian frame, where the metric deviates from
the Minkowski metric by simple corrections, and where the equations of
motion are slightly modified Newtonian equations \citep{Soffel00}.

Are these coordinate systems \emph{methodologically preferred} because of
their convenience?  If so, can they confer identity to the
point-events of space-time?  Both questions deserve some
investigation; however, we should note that they do not refer directly
to the philosophical analysis of general relativity in the generic
case, but rather in the case of a specific solution (our universe).
So we should be cautious when we discuss the connection between the
physical individuation of points (as we have outlined it) and the
\emph{theory of measurement} in general relativity, with its many
real-world applications (such as time transport, geographic
positioning and solar-system navigation).  The \emph{practice} (but
not the theory) of general-relativistic measurements is necessarily a
consequence of the particular solution of the Einstein equations that
we happen to inhabit.\footnote{On the contrary, the physical
individuation of points events by the analysis of the local metric
fingerprint would be very relevant to orientation and navigation in a
hypothetical world that is devoid of matter, and where gravity is very
strong and unpredictable.}

Still, we wish to draw a scenario of how the physical individuation of
points could be implemented (in principle) as an experimental setup
and protocol for positioning and orientation.  This construction,
which could also be discussed more abstractly as a system of
axioms\footnote{We owe the classical paper on the axiomatics of
general relativity to Ehlers, Pirani and Schild
\citeyearpar{Ehlerspiranischild00}, who start out by defining basic
objects such as light rays, freely falling test particles, standard
clocks, and so on. In their scheme, light-ranging measurements are
then used to reveal the \emph{conformal} structure of space-time, while
the free fall of test bodies is used to map out the \emph{projective}
structure. Under an axiom of compatibility [well corroborated by
experiment; see \citet{Perlick00}] these two classes of observations
determine completely the structure of space-time.

We note here that both the Ehlers--Pirani--Schild axiomatics
(based on idealized primitive physical objects and operations) and
our discussion of coordinate systems and metric field measurements
in terms of technological instruments (GPS satellites) imply that
the \emph{coordination} of the mathematical theory of general
relativity to the physical quantities defined operationally cannot
be excised from the wider context of a comprehensive theory of
physical reality, where the idealized primitive objects and
operations of Ehlers, Pirani and Schild are, in essence,
implemented by our technological instruments.} for the empirical
foundation of general relativity, \emph{closes the coordinative
circuit} that joins the mathematical formulation of general
relativity (and in particular of the Hamiltonian initial-value
problem) to the practice of general-relativistic measurement, and
to the physical individuation of space-time points. Three steps
are necessary.
\begin{enumerate}
\item We define a \emph{radar-gauge system of coordinates} in a finite
four-dimensional volume, by means of a network of artificial
satellites similar to the Global Positioning System
\citep{Ashbyspilker00}.  The GPS is a constellation of 24 satellites
on quasicircular 20-km-high orbits around the Earth; each GPS
satellite carries an atomic clock accurate to the nanosecond, and
continuously broadcasts its own position and time,\footnote{More
precisely, the clocks on the satellites are biased to yield the
international standard time; that is, the proper time elapsed on the
\emph{geoid}, the surface of constant effective gravitational
potential that sits very close to the surface of the Earth (at sea
level).} as computed within an accurate model of its motion in the
gravitational field of the Earth. By comparing the signals received
from four satellites at a given instant of time
(\emph{pseudo-ranging}), the \emph{GPS receivers} on the surface of
the Earth are able to determine their radar distance from the
satellites, and therefore to compute their own latitude, longitude,
and altitude with a precision of a few tens of meters, and to track
the international standard time with a maximum error of a few
nanoseconds.

The GPS receivers are able to determine their actual position (that
is, the set of their four post--Newtonian, geocentric coordinates,
with the time coordinate rescaled to the international standard time),
because the entire GPS system is predicated on the advance knowledge
of the gravitational field of the Earth and of the trajectories of the
satellites, which in turn allows the \emph{coordinate} synchronization
of the satellite clocks to post--Newtonian time. If, as in our case,
the geometry of space-time and the motion of the satellites are not
known in advance, it would be still possible for the receivers to
obtain four, as it were, \emph{conventional} coordinates by operating
a full-ranging protocol (involving bidirectional communication) to
four super-GPS satellites that broadcast the time of their standard,
unsynchronized clocks. The problem of patching the coordinates
obtained from different four-tuples of satellites is analog to
deriving the coordinate transformations between overlapping patches
within an atlas of a differential manifold, and it should be tractable
by maintaining full-ranging communication between the satellites
themselves.

Summarizing, our super-GPS constellation provides a radar-gauge
system of coordinates (without any direct metrical significance)
for all the point-events within a finite region of
space-time\footnote{Within the Ehlers--Pirani--Schild axiomatics,
this corresponds to determining the \emph{conformal structure} of
space-time.}:
\begin{equation}
\sigma_R^A \equiv (\tau_R,\sigma_R^a); \quad \tau_R = 0 \;
\mathrm{defines} \; \Sigma_{\tau_R}.
\end{equation}
\item By means of repeated measurements of the motion of four test
particles\footnote{For vacuum gravitational fields. Six test particles
are needed in general space-times.} (see Ciufolini and Wheeler \citeyear{Ciufoliniwheeler00}, pp.\ 34--36; see also Rovelli \citeyear{Rovelli03}) and gyroscopes (to measure $N^A$!), with
technologies similar to the Gravity Probe B space mission \citep{GPB},
suitable spacecraft could then \emph{measure the components of the
four-metric with respect to the radar-gauge coordinates},
\begin{equation}
{}^4 g_{R(A,B)} (\tau_R,\sigma_R^a),
\end{equation}
and by measuring the spatial and temporal variation of ${}^4 g$,
we could then compute (in principle) the components of the Weyl tensor, and
\emph{the Weyl invariant scalars}.\footnote{Within the
Ehlers--Pirani--Schild axiomatics, this corresponds to determining
the \emph{projective structure} of space-time.}
\item By steps 1 and 2, we have obtained a slicing of space-time
into surfaces $\Sigma_{\tau_R}$, and a set of coordinates $\sigma^a$
on the surfaces, both defined \emph{operationally}; furthermore, we
have determined the components of the metric and the local value of
the Weyl scalars with respect to the $\sigma^A$. We can then solve (in principle) for
the functions $\hat{Z}^{[A]}$ that reproduce the radar-gauge
coordinates as \emph{radar-gauge intrinsic coordinates},
\begin{equation}
\sigma_R^A =
\hat{Z}^{[A]}\bigl[w_T[g(\sigma_R^B),\pi(\sigma_R^C)]\bigr].
\end{equation}
Finally, we can impose the gauge fixing that enforces this
particular system of intrinsic coordinates,
\begin{equation}
\chi^A \equiv \sigma^A - \hat{Z}^{[A]}
\bigl[w_T[g(\sigma^B),\pi(\sigma^C)]\bigr] \approx 0;
\end{equation}
at the end of the canonical reduction procedure, we obtain the
structure of the Dirac observables $q^r$, $p_s$ as nonlocal
functionals of $g$ and $\pi$, and we reconstruct the intrinsic
coordinates as functions of the Dirac observables in each
point-event of space-time:
\begin{equation}
\sigma_R^A =
\hat{Z}^{[A]}\bigl[w_T[q^r(\sigma_R^B),p_s(\sigma_R^C)]\bigr].
\end{equation}
Thus, the radar-gauge coordinates are legitimized as intrinsic
coordinates that, because of their well-defined dependence on the
Dirac observables, can endow the point-events of space-time with
physical individuality.  Of course, the particular form of this
dependence, and the particular \emph{presentation} of the \emph{true}
degrees of freedom of the gravitational field is gauge dependent.
\end{enumerate}

\noindent This procedure closes the \emph{coordinative circuit} of general relativity, linking individuation to experimentation.

\section{Conclusion: finding the last remnant of physical objectivity}
\label{sec:conclusion}

From the point of view of the constrained Hamiltonian formalism, general relativity is a gauge theory like any other; however, it is radically different from the physical point of view.
In addition to creating the distinction between what is observable\footnote{In the Dirac or Bergmann sense.} and
what is not, the gauge freedom of general relativity is unavoidably entangled with the
definition--constitution of the very \emph{stage}, space-time, where the \emph{play} of physics is enacted.  In other words, the gauge mechanism has the double role of making the dynamics unique (as in all gauge theories), and of fixing the spatio-temporal reference background \emph{at the mathematical level}.

In gauge theories such as electromagnetism, we can rely from the beginning on empirically validated, gauge-invariant dynamical equations for the \emph{local} fields. This is not the case for general relativity: in order to get dynamical equations for the basic
field in a \emph{local} form, we must pay the price of general covariance, which weakens the objectivity that the
spatio-temporal description could have had \emph{a priori}.
Recalling the definition of gauge theory given by Henneaux and Teitelboim (see the beginning of Sec.\ \ref{sec:congen}), we could say that the introduction of extra variables does make the mathematical description of general relativity more transparent, but it also makes its physical interpretation more obscure and intriguing, at least at first sight.

By now, it should be clear that the Hole Argument has nothing to do with the alleged \emph{indeterminism} of general relativity as a dynamical theory. In our discussion of the initial-value problem within the Hamiltonian framework we have shown that, \emph{on shell}, a complete gauge-fixing (which could in theory concern the whole space-time) is equivalent to the choice of an atlas of coordinate charts on the space-time manifold, and in particular \emph{within the Hole}.
At the same time, we have seen that the active diffeomorphisms of the manifold can be interpreted as passive Hamiltonian gauge transformations.
Because the gauge must be fixed \emph{before} the initial-value problem can be solved to obtain a solution (outside and inside the Hole), it makes little sense to apply active diffeomorphisms to an already generated solution to obtain an allegedly ``different'' space-time. Conversely, it should be possible to generate these ``different'' solutions by appropriate choices of the initial gauge fixing.

In addition, we have established that within the Hamiltonian framework we can use a gauge-fixing procedure based on the Bergmann--Komar intrinsic coordinates to turn the \emph{primary} mathematical individuation of manifold \emph{points} into a \emph{physical} individuation of \emph{point-events} that is directly associated with the value of the gravitational degrees of freedom (Dirac observables).
The price to pay is the breaking of general covariance.
General covariance thus represents a horizon of \emph{a priori} possibilities for the physical constitution of the space-time, possibilities that must be actualized within any given solution of the dynamical equations.  What here we called \emph{physical constitution} embodies at the same time the \emph{chrono-geometrical}, the \emph{gravitational}, and the \emph{causal} properties of the space-time stage.

We have shown that this conceptual physical individuation can be implemented (at least in principle) with a well-defined empirical procedure that closes the \emph{coordinative circuit}. We believe that these results cast some light over the \emph{intrinsic structure} of the general relativistic space-time that had disappeared within Leibniz equivalence and that was the object of Michael Friedman's non-trivial question.

In 1972, Bergmann and Komar wrote \citep{Bergmannkomar01}:
\begin{quotation}
\noindent [...] in general relativity the identity of a world point is not preserved under the theory's widest invariance group. This assertion forms the basis for the conjecture that some physical theory of the future may teach us how to dispense with world points as the ultimate constituents of space-time altogether.
\end{quotation}
Indeed, would it be possible to build a fundamental theory that is grounded in the reduced phase space parametrized by the Dirac observables? This would be an abstract and highly nonlocal theory of gravitation that would admit an infinity of gauge-related, spatio-temporally local realizations.
From the mathematical point of view, however, this theory would be just an especially perspicuous instantiation of the relation between canonical structure and locality that pervades contemporary theoretical physics nearly everywhere.

On the other hand, beyond the mathematical transparency and the latitude of choices guaranteed by general covariance, we need to remember that \emph{local} spatio-temporal realizations of the abstract theory would still be needed for implementation of measurements in practice; conversely, any real-world experimental setting entails the choice of a definite \emph{local realization}, with a corresponding gauge fixing that breaks general covariance.

Can this basic freedom in the choice of the \emph{local
realizations} be equated with a ``taking away from space and time
the last remnant of physical objectivity,'' as Einstein suggested?
We believe that if we strip the physical situation from
Einstein's ``spatial obsession'' about \emph{realism as locality (and separability)}, a
significant kind of spatio-temporal objectivity survives.
It is true that the \emph{functional dependence} of the Dirac observables upon the spatio-temporal coordinates depends on the particular choice of the latter (or equivalently, of the gauge); yet, there is no \emph{a priori} \emph{physical} individuation of the points independently of the metric field, so we cannot say that the physical-individuation procedures corresponding to different gauges individuate physical point-events that are \emph{really} different. Given the conventional nature of the primary mathematical individuation of manifold points through $n$-tuples of real numbers, we could say instead that the \emph{real point-events} are constituted by the nonlocal values of gravitational degrees of freedom, while the underlying point structure of the mathematical manifold may be changed at will.

In conclusion, we have presented evidence that the non--point-like texture encoded in the Dirac observables allows a conception of space-time that is a \emph{new kind of structuralism}, in the tradition of Newton's \emph{De Gravitatione}, only much richer. This is even more evident in the case of general relativity with matter, where we have Dirac observables both for the gravitational field and for the matter fields, and where the former are modified in their functional form by the presence of matter. Since the gravitational Dirac observables will still provide the individuating fields for point-events (according to the conceptual structure discussed in this paper), \emph{matter will come to influence the very individuation of points}. Thus, our structuralist view is richer also in a deeper sense, because it includes elements in the tradition of both absolutism (space has an autonomous existence independently of other bodies or matter fields) \emph{and} relationism (the nature of space depends on the relations between bodies, or space has no reality independently of the fields it contains).

A future direction of investigation is the following: looking at the Bergmann--Komar \emph{intrinsic
components} of the metric [see Eq.\ \eqref{eq:deceiving}], and calculating the Dirac brackets of the Weyl scalars, it might be possible to define four diffeomorphically invariant and canonically conjugated
variables that are also Dirac observables \emph{on shell}. This achievement would unify the general-covariant and the Dirac--Bergmann--Komar notion of observable, and would provide explicit evidence for the
objectivity of point-event individuation.
Finally, the procedure of individuation that we have outlined transfers, as it were, the noncommutative Poisson--Dirac structure of the Dirac observables onto the individuated point-events; the physical implications of this circumstance might deserve some attention in view of the quantization of general relativity.


\begin{thebibliography}{99}
%
\bibitem[Abramowicz(1993)]{Abramowicz00} Abramowicz, M. A. (1993),
``Inertial Forces in General Relativity'', in \emph{The Renaissance of
General Relativity and Cosmology}, G. Ellis, A. Lanza and J. Miller
(eds.), (Cambridge University, Cambridge), pp. 40--58.
%
\bibitem[Abramowicz, Nurowski and Wex(1993)]{Abramowicznurowskiwex00} Abramowicz, M. A., Nurowski, P., and
Wex, N. (1993), ``Covariant Definition of Inertial Forces'',
\emph{Class. Quantum Grav.} \textbf{10}, L183--L186.
%
\bibitem[Arnowitt, Deser and Misner(1962)]{Adm00} Arnowitt, R., Deser, S., and Misner, C. W. (1962),
``The dynamics of general relativity'', in \emph{Gravitation: an
introduction to current research}, L. Witten (ed.) (Wiley, New York,
1962), pp. 227--265.
%
\bibitem[Ashby and Spilker(1995)]{Ashbyspilker00} Ashby, N., and Spilker, J. J. (1995),
``Introduction to Relativistic Effects on the Global Positioning
System'', in \emph{Global Positioning System: Theory and
Applications}, Vol. 1, B. W. Parkinson and J. J.  Spilker (eds.)
(American Institute of Aeronautics and Astronautics), 623--697.
%
\bibitem[Bartels(1994)]{Bartels00} Bartels, A. (1994), ``What is space-time if not a
Substance? Conclusions from the New Leibnizian Argument'', in
\emph{Semantical Aspects of space-time Theories}, U. Mayer and
H.--J. Schmidt (eds.) (B. I. Wissenshaftverlag, Mannheim), pp. 41--51.
%
\bibitem[Belot(1995)]{Belot00} Belot, G. (1995), ``Indeterminism and Ontology'',
\emph{International Studies in the Philosophy of Science} \textbf{9},
8--101.
%
\bibitem[Belot(1996)]{Belot01} Belot, G. (1996), ``Why General Relativity does Need
an Interpretation'', in \emph{Philosophy of Science Suppl.}
\textbf{63}, S80--S88.
%
\bibitem[Bergmann(1960)]{Bergmann00} Bergmann, P. G. (1960), ``Observables in General
Relativity'', \emph{Rev. Mod. Phys.} \textbf{33}, 510--514.
%
\bibitem[Bergmann(1962)]{Bergmann02} Bergmann, P. G. (1962), ``The General Theory of
Relativity'', in \emph{Handbuch der Physik}, Vol. IV, \emph{Principles
of Electrodynamics and Relativity}, S. Flugge (ed.) (Springer-Verlag,
Berlin), pp. 247--272.
%
\bibitem[Bergmann(1971)]{Bergmann01} Bergmann, P. G. (1971), ``Hamilton--Jacobi Theory
with Mixed Constraints'', N. Y. A. S., Scientific Award for 1970 (New
York Academy of Sciences, New York).
%
\bibitem[Bergmann(1977)]{Bergmann03} Bergmann, P. G. (1977), ``Geometry and
Observables'', in \emph{Foundations of Space-Time Theories, Minnesota
Studies in the Philosophy of Science}, Vol. VIII, J. S. Earman,
C. N. Glymour, and J. Stachel (eds.) (University of Minnesota,
Minneapolis), pp. 275--280.
%
\bibitem[Bergmann and Komar(1960)]{Bergmannkomar00} Bergmann, P. G., and Komar, A. (1960),
``Poisson Brackets between Locally Defined Observables in General
Relativity'', \emph{Phys. Rev. Lett.} {\textbf 4}, 432--433.
%
\bibitem[Bergmann and Komar(1972)]{Bergmannkomar01} Bergmann, P. G., and Komar, A. (1972), ``The
Coordinate Group Symmetries of General Relativity'',
\emph{Int. J. Theor. Phys.} {\textbf 5}, 15--28.
%
\bibitem[Brighouse(1994)]{Brighouse00} Brighouse, C. (1994), ``space-time and Holes'', in \emph{PSA 1994}, \textbf{1}, D. Hull, M. Forbes and R. M. Burian (eds.), pp. 117--125.
%
\bibitem[Butterfield(1984)]{Butterfield03} Butterfield, J. (1984), ``Relationism and
Possible Worlds'', \emph{British Journal for the Philosophy of
Science} \textbf{35}, 1--13.
%
\bibitem[Butterfield(1987)]{Butterfield00} Butterfield, J. (1987), ``Substantivalism and
Determinism'', \emph{International Studies in the Philosophy of
Science} \textbf{2}, 10--31.
%
\bibitem[Butterfield(1988)]{Butterfield01} Butterfield, J. (1988), ``Albert Einstein meets David Lewis'', in
\emph{PSA 1988}, \textbf{2}, A. Fine and J. Leplin (eds.), pp.56--64.
%
\bibitem[Butterfield(1989)]{Butterfield02} Butterfield, J. (1989), ``The Hole Truth'',
\emph{British Journal for the Philosophy of Science} \textbf{40},
1--28.
%
\bibitem[Chuang(1996a)]{Chuang00} Chuang, L. (1996a), ``Realism and space-time: Of
Arguments Against Metaphysical Realism and Manifold Realism'',
\emph{Philosophia Naturalis} \textbf{33}, 24--63.
%
\bibitem[Chuang(1996b)]{Chuang01} Chuang, L. (1996b), ``Gauge Invariance, Cauchy
Problem, Indeterminism, and Symmetry Breaking'', in \emph{PSA 1996},
\textbf{63}, S71--S79.
%
\bibitem[Ciufolini and Wheeler(1995)]{Ciufoliniwheeler00} Ciufolini, I. and Wheeler, J. A. (1995), \emph{Gravitation and Inertia} (Princeton University Press, Princeton), pp. 34--36.
%
\bibitem[Disalle(1994)]{Disalle00} Disalle, R. (1994), ``On Dynamics,
Indiscernibility, and space-time Ontology'', \emph{British Journal for
the Philosophy of Science} \textbf{45}, 265--287.
%
\bibitem[Dorato(2000)]{Dorato00} Dorato, M. (2000), ``Substantivalism, Relationism,
and Structural Space-time Realism'', \emph{Foundations of Physics}
\textbf{30}, 1605--1628.
%
\bibitem[Earman(1979)]{Earman00} Earman, J. (1979), ``Was Leibniz a Relationist?'' in \emph{Studies in Metaphysics},
Midwest Studies in Philosophy, Vol. 4, P. French and H. Wettstein (eds.) (University of Minnesota Press, Minneapolis), pp. 263--276.
%
\bibitem[Earman and Norton(1987)]{Earmannorton00} Earman, J., and Norton, J. (1987),
``What Price space-time Substantivalism? The Hole Story'',
\emph{British Journal for the Philosophy of Science} \textbf{38},
515--525.
%
\bibitem[Ehlers, Pirani and Schild(1972)]{Ehlerspiranischild00} Ehlers,
J., Pirani, F. A. E., and Schild, A. (1972), ``The Geometry of Free
Fall and Light Propagation'', in \emph{General Relativity},
L. O'Raifeartaigh (ed.), (Clarendon, Oxford), pp. 63--84.
%
\bibitem[Einstein(1914)]{Einstein01} Einstein, A. (1914) ``Die formale Grundlage der
allgemeinen Relativit\"atstheorie'', in \emph{Preuss. Akad. der
Wiss. Sitz.}, 1030--1085.
%
\bibitem[Einstein(1916)]{Einstein00} Einstein, A. (1916), ``Die Grundlage der allgemeinen Relativit\"atstheorie'', \emph{Annalen der Physik} \textbf{49}, 769--822; translation by W.  Perrett and G. B. Jeffrey (1952), ``The Foundation of the General Theory of Relativity'', in \emph{The Principle of Relativity} (Dover, New York), pp. 117--118.
%
\bibitem[Fine(1984)]{Fine00} Fine, A. (1984), ``The Natural Ontological
Attitude'', in \emph{Scientific Realism}, J. Leplin (ed.) (University
of California Press, Berkeley), pp. 253--316.
%
\bibitem[Friedman(1983)]{Friedman00} Friedman, M. (1983), \emph{Foundations of Space-Time Theories} (Princeton University, Princeton), pp. 221--223.
%
\bibitem[Friedman(1984)]{Friedman02} Friedman, M. (1984) ``Roberto Torretti, \emph{Relativity and Geometry}'', critical review, \emph{No\^us} \textbf{18}, 653--664.
%
\bibitem[Friedman(2001)]{Friedman01} Friedman, M. (2001),
``Geometry as a Branch of Physics: Background and Context for Einstein's `Geometry and
Experience'{''}, in \emph{Reading Natural Philosophy: Essays in the
History and Philosophy of Science and Mathematics to Honor Howard
Stein on His $70^{th}$ Birthday}, D. Malament (ed.) (Open Court,
Chicago).
%
\bibitem[GPB()]{GPB} Gravity Probe B website: \url{http://einstein.stanford.edu}.
%
\bibitem[G\'eh\'enieau and Debever(1956)]{Geheniaudebever00} G\'eh\'eniau, J. and Debever, R. (1956),
``Les quatorze invariants de courbure de l'espace riemannien \`a
quatre dimensions'', in \emph{Proceedings of the Jubilee of Relativity
Theory}, Bern, July 11-16, 1955, A. Mercier and M. Kervaire (eds.)
(Birkh\"auser Verlag, Basel), pp. 101--105.
%
\bibitem[Hall and Hall(1962)]{HallandHall} Hall, A. R., and Hall, M. B. (eds.) (1962),
\emph{De Gravitatione et Aequipondio Fluidorum, Unpublished Scientific
Papers of Isaac Newton. A Selection from the Portsmouth Collection in
the University Library} (Cambridge University, Cambridge).
%
\bibitem[Henneaux and Teitelboim(1992)]{Henneauxteitelboim00} Henneaux, M. and Teitelboim, C. (1992),
\emph{Quantization of Gauge Systems} (Princeton University,
Princeton).
%
\bibitem[Hilbert(1917)]{Hilbert00} Hilbert, D. (1917), ``Die
Grundlagen der Physik. (Zweite Mitteilung)'', \emph{Nachrichten
von der K\"oniglichen Gesellschaft der Wissenschaften zu
G\"ottingen, Mathematisch-physikalische Klasse}, 53--76.
%
\bibitem[Hofer(1996)]{Hofer00} Hofer, C. (1996), ``The Metaphysics of Space-Time
Substantivalism'', \emph{Journal of Philosophy} \textbf{93}, 5--27.
%
\bibitem[Hofer(1998)]{Hofer01} Hofer, C. (1998), ``Absolute versus Relational
space-time: For Better or Worse, the Debate Goes On'', \emph{British
Journal for the Philosophy of Science} \textbf{49}, 451--467.
%
\bibitem[Hofer and Cartwright(1993)]{Hofercartwright00} Hofer, C. and Cartwright, N. (1993),
``Substantivalism and the Hole Argument'', in \emph{Philosophical
Problems of the Internal and External Worlds, Essays on the Philosophy
of Adolf Gruenbaum}, J. Earman, I. Janis, G. J. Massey and N. Rescher
(eds.) (University of Pittsburgh Press, Pittsburgh), pp. 23--43.
%
\bibitem[Howard and Norton(1993)]{Howardnorton00} Howard, D. and Norton, J. (1993) ``Out of
the Labyrinth? Einstein, Hertz, and G\"ottingen Answer to the
Hole Argument'', in \emph{The Attraction of Gravitation: New Studies in the History of General
Relativity}, Einstein Studies, Vol. 5, J. Earman, M. Jansen, J. Norton, (eds.) (Birkh\"auser, Boston), pp. 30--62.
%
\bibitem[Komar(1955)]{Komar00} Komar, A. (1955), ``Degenerate Scalar Invariants and
the Groups of Motion of a Riemann Space'',
\emph{Proc. Natl. Acad. Sci.} \textbf{41}, 758--762.
%
\bibitem[Komar(1958)]{Komar01} Komar, A. (1958), ``Construction of a Complete Set of
Independent Observables in the General Theory of Relativity'',
\emph{Phys. Rev.} \textbf{111}, 1182--1187.
%
\bibitem[Leeds(1995)]{Leeds00} Leeds, S. (1995), ``Holes and Determinism: Another
Look'', \emph{Philosophy of Science} \textbf{62}, 425--478.
%
\bibitem[Lusanna(2000)]{Lusanna01} Lusanna, L. (2000), ``Towards a Unified
Description of the Four Interactions in Terms of Dirac--Bergmann
Observables'', in \emph{Quantum Field Theory}, A. N. Mitra (ed.)
(Indian National Science Academy), pp. 490--518. e-print
\texttt{hep-th/9907081}.
%
\bibitem[Lusanna(2001)]{Lusanna00} Lusanna, L. (2001), ``The Rest-Frame Instant Form
of Metric Gravity'', \emph{Gen. Rel. Grav.} \textbf{33},
1579--1696. e-print \texttt{gr-qc/0102074}.
%
\bibitem[Lusanna and Pauri(2002)]{Lusannapauri00} Lusanna L., and Pauri, M. (2002),
``Individuating space-time Points by Structure: The Role of
Gravitational and Gauge Degrees of Freedom in General Relativity'', in
preparation.
%
\bibitem[Maudlin(1988)]{Maudlin00} Maudlin, T. (1988), ``The Essence of Space-Time'',
in \emph{PSA 1988}, \textbf{2}, pp. 82--91.
%
\bibitem[Maudlin(1990)]{Maudlin01} Maudlin, T. (1990), ``Substances and space-times:
What Aristotle Would Have Said to Einstein'', \emph{Studies in the
History and Philosophy of Science} \textbf{21}, 531--61.
%
\bibitem[Norton(1984)]{Norton06} Norton, J. (1984), ``How Einstein found his Field
Equations: 1912--1915'', \emph{Historical Studies in the Physical
Sciences}, \textbf{14}, 252-316; reprinted in \emph{Einstein and the History of General Relativity}, Einstein
Studies, Vol. 1, D. Howard and J. Stachel
(eds.) (Birkh\"auser, Boston), pp. 101--159.
%
\bibitem[Norton(1987)]{Norton00} Norton, J. (1987),
``Einstein, the Hole Argument and the Reality of Space'', in
\emph{Measurement, Realism and Objectivity}, J. Forge (ed.)
(Reidel, Dordrecht), pp. 153--18.
%
\bibitem[Norton(1988)]{Norton05} Norton, J. (1988), ``The Hole Argument'', \emph{PSA 1988}, Vol. 2, 56--64.
%
\bibitem[Norton(1989)]{Norton07} Norton, J. (1989), ``Coordinate and
Covariance: Einstein's View of Space-Time and the Modern View'',
\emph{Foundations of Physics}, \textbf{19}, 1215--1263.
%
\bibitem[Norton(1992)]{Norton01} Norton, J. (1992), ``The Physical Content of General
Covariance'', in \emph{Studies in the History of General
Relativity}, Einstein Studies, Vol. 3, J. Eisenstaedt, and A. Kox
(eds.) (Birkh\"auser, Boston), pp. 281--315.
%
\bibitem[Norton(1993)]{Norton02} Norton, J. (1993), ``General Covariance and the
Foundations of General Relativity: Eight Decades of Dispute'',
\emph{Rep. Prog. Phys.} \textbf{56}, 791--858.
%
\bibitem[Norton(2001)]{Norton04} Norton, J. (2001), ``General Covariance, Gauge Theories and the Kretschmann Objection'', e-print
\url{PITT-PHIL-SCI00000380}, August 2001.
%
\bibitem[Norton(2002)]{Norton03} Norton, J. (2002), ``Einstein's
Triumph over the Space-Time Coordinate System'', \emph{Di\'alogos} \textbf{79}, 253--262.
%
\bibitem[Pauri(1996)]{Pauri00} Pauri, M. (1996), ``Realt\`a e Oggettivit\`a'', in \emph{L'Oggettivit\`a nella Conoscenza Scientifica} (F. Angeli, Brescia), pp. 79--112.
%
\bibitem[Perlick(1994)]{Perlick00} Perlick, V. (1994), ``Characterization of Standard
Clocks in General Relativity'', in \emph{Semantical Aspects of
space-time Theories}, U. Mayer and H.--J. Schmidt (eds.)
(B. I. Wissenshaftverlag, Mannheim), pp. 169--179.
%
\bibitem[Rovelli(1991)]{Rovelli00} Rovelli, C. (1991), ``What is Observable in
Classical and Quantum Gravity?'', \emph{Class. Quantum Grav.}
\textbf{8}, 297--316.
%
\bibitem[Rovelli(1997)]{Rovelli01} Rovelli, C. (1997), ``Halfway Through the Woods:
Contemporary Research on Space and Time'', in \emph{The Cosmos of
Science}, J. Earman and J. D. Norton (eds.) (University of Pittsburgh
and Universit\"ats Kostanz Verlag, Pittsburgh-Konstanz), pp. 180--223.
%
\bibitem[Rovelli(1999)]{Rovelli02} Rovelli, C. (1999), ``Quantum space-time: What Do
We Know?'', in \emph{Physics meets Philosophy at the Planck Scale},
C. Callender, N. Hugget (eds.) (Cambridge University, Cambridge),
pp. 153--201.
%
\bibitem[Rovelli(2001)]{Rovelli03} Rovelli, C. (2001), ``GPS observables in general relativity'', e-print \url{gr-qc/0110003}, October 2001.

\bibitem[Rynasiewicz(1992)]{Rynasiewicz00} Rynasiewicz, R. (1992), ``Rings, Holes and
Substantivalism: On the Program of Leibniz Algebras'',
\emph{Philosophy of Science} \textbf{59}, 572--89.
%
\bibitem[Rynasiewicz(1994)]{Rynasiewicz01} Rynasiewicz, R. (1994), ``The Lessons of the
Hole Argument'', \emph{British Journal for the Philosophy of Science}
\textbf{45}, 407--436.
%
\bibitem[Rynasiewicz(1996a)]{Rynasiewicz02} Rynasiewicz, R. (1996a), ``Is There a Syntactic Solution to the Hole Problem?'', in \emph{PSA 1996}, \textbf{63}, S55--S62.
%
\bibitem[Rynasiewicz(1996b)]{Rynasiewicz03} Rynasiewicz, R. (1996b), ``Absolute versus
Relational Space-Time: An Outmoded Debate?'', \emph{Journal of
Philosophy} \textbf{43}, 279--306.
%
\bibitem[Saunders(2001)]{Saunders00} Saunders, S. (2001), ``Indiscernibles, General
Covariance, and Other Symmetries'', e-print
\texttt{PITT-PHIL-SCI00000459}, October 2001.
%
\bibitem[Schlick(1917)]{Schlick00} Schlick, M. (1917), \emph{Raum, Zeit in der
gegenw\"artigen Physik} (Springer, Berlin); reprinted in H. Mulder and
B. van der Velde-Schlick, (eds.) (1978), \emph{Moritz Schlick:
Philosophical Papers}, Vol. 1 (Reidel, Dordrecht), p. 260.
%
\bibitem[Shanmugadhasan(1973)]{Shanmugadhasan00} Shanmugadhasan, S. (1973),
``Canonical Formalism for Degenerate Lagrangians'',
\emph{Journ. Math. Phys.} \textbf{14}, 677--687.
%
\bibitem[Soffel(1989)]{Soffel00} Soffel, M. H. (1989), \emph{Relativity in
Astronometry, Celestial Mechanics and Geodesy} (Springer-Verlag,
Berlin).
%
\bibitem[Stachel(1980)]{Stachel00} Stachel, J. (1980), ``Einstein's Search for General
Covariance, 1912--1915'', paper read at the Ninth International
Conference on General Relativity and Gravitation, Jena; published
in \emph{Einstein and the History of General Relativity}, Einstein
Studies, Vol. 1, D. Howard and J. Stachel (eds.) (Birkh\"auser,
Boston 1986), pp. 63--100.
%
\bibitem[Stachel(1986a)]{Stachel04} Stachel, J. (1986a), ``How Einstein Discovered
General Relativity: A Historical Tale with Some Contemporary
Morals'', in \emph{General Relativity and Gravitation}, M. A. H.
MacCallum (ed.) (Cambridge University Press, Cambridge),
pp. 200--208.
%
\bibitem[Stachel(1986b)]{Stachel01} Stachel, J. (1986b), ``What can a Physicist Learn
from the Discovery of General Relativity?'', \emph{Proceedings of
the Fourth Marcel Grossmann Meeting on Recent Developments in
General Relativity, Rome, 1--21 June, 1985}, R. Ruffini (ed.)
(North-Holland, Amsterdam).
%
\bibitem[Stachel(1992)]{Stachel05} Stachel, J. (1992), ``The Cauchy Problem in General Relativity -- The Early Years'', in \emph{Historical
Studies in General Relativity}, Einstein Studies, Vol. 3, J. Eisenstaedt and A. J. Kox (eds.) (Birkh\"auser, Boston), pp.
407--418.
%
\bibitem[Stachel(1993)]{Stachel02} Stachel, J. (1993), ``The Meaning of General
Covariance -- The Hole Story'', in \emph{Philosophical Problems of the
Internal and External Worlds, Essays on the Philosophy of Adolf
Gruenbaum}, J. Earman, I. Janis, G. J. Massey and N. Rescher (eds.)
(University of Pittsburgh Press, Pittsburgh), pp. 129--160.
%
\bibitem[Stachel(1999)]{Stachel03} Stachel, J. (1999), ``Space-Time Structures:
What's The Point?'', Talk delivered at the Minnowbrook Symposium, May
28--31, 1999.
%
\bibitem[Stein(1977)]{Stein00} Stein, H. (1977), ``On Space-Time and Ontology:
Extract from a Letter to Adolf Gr\"unbaum'', in \emph{Foundations of
Space-Time Theories}, Minnesota Studies in the Philosophy of Science,
Vol. VIII, J. S. Earman, C. N. Glymour, and J. J. Stachel (eds.)
(University of Minnesota, Minneapolis), pp. 374--402.
%
\bibitem[Sundermayer(1982)]{Sundermayer00} Sundermayer, K. (1982), \emph{Constrained
dynamics} (Springer--Verlag, Berlin).
%
\bibitem[Teller(1991)]{Teller00} Teller, P. (1991), ``Substances, Relations and
Arguments About the Nature of space-time'', \emph{The Philosophical
Review} \textbf{C}, 3, 363--97.
%
\bibitem[Torretti(1987)]{Torretti00} Torretti, R. (1987), \emph{Relativity and
Geometry}, (Dover, New York), pp. 167--168.
%
\bibitem[Torretti(1999)]{Torretti01} Torretti, R. (1999), \emph{The Philosophy of
Physics} (Cambridge University Press, Cambridge), p. 297.
%
\bibitem[Wald(1984)]{Wald00} Wald, R. M. (1984), \emph{General Relativity},
(University of Chicago, Chicago), pp. 438--439.
%
\bibitem[Weyl(1946)]{Weyl00} Weyl, H. (1946), ``Groups, Klein's Erlangen
Program. Quantities'', ch. I, sec. 4 of \emph{The Classical Groups,
Their Invariants and Representations}, 2nd ed., (Princeton University,
Princeton), pp. 13--23.
%
\bibitem[Wilson(1993)]{Wilson00} Wilson, M. (1993), ``There's a Hole and a Bucket,
Dear Leibniz'', in \emph{Philosophy of Science}, P. A. French,
T. E. Uehling and H. K. Wettstein (eds.) (University of Notre Dame,
Notre Dame).
%
\end{thebibliography}
\end{document}